\documentclass[amsmath,amssymb,aps,floatfix,jcp,superscriptaddress,preprint]{revtex4-1}

\usepackage[utf8]{inputenc} 
\usepackage[T1]{fontenc}    
\usepackage{amsfonts}       
\usepackage{booktabs}       
\usepackage{hyperref}       
\usepackage{url}            
\usepackage{microtype}      
\usepackage{nicefrac}       
\usepackage{paralist}       
\usepackage{graphicx}
\usepackage{float}
\usepackage{wrapfig}
\usepackage[dvipsnames]{xcolor}
\usepackage{ulem}
\usepackage{tikz}
\usetikzlibrary{arrows,shapes,angles,quotes}
\usetikzlibrary{decorations.pathreplacing}
\usepackage{fullpage}

\usepackage{lmodern}
\usepackage{ulem}

\newcommand*{\vcenteredhbox}[1]{\begin{tabular}{@{}l@{}}#1\end{tabular}}

\begin{document}
	
	\title{Machine learning interatomic potentials for aluminium: application to solidification phenomena}
	\author{Noel Jakse}
	\affiliation{Université Grenoble Alpes, CNRS, Grenoble INP, SIMaP
		F-38000 Grenoble, France}
	
	\author{Johannes Sandberg}
	\affiliation{Université Grenoble Alpes, CNRS, Grenoble INP, SIMaP
		F-38000 Grenoble, France}
	\affiliation{Institut für Materialphysik im Weltraum, Deutsches Zentrum für Luft- und Raumfahrt (DLR), 51170 	
		Köln, Germany }
	\affiliation{Department of Physics, Heinrich-Heine-Universität Düsseldorf, Universitätsstraße 1, 40225 Düsseldorf, Germany}
	\author{Leon F. Granz}
	\affiliation{Institut für Materialphysik im Weltraum, Deutsches Zentrum für Luft- und Raumfahrt (DLR), 51170 	
		Köln, Germany }
	\affiliation{Department of Physics, Heinrich-Heine-Universität Düsseldorf, Universitätsstraße 1, 40225 Düsseldorf, Germany}
	
	\author{Anthony Saliou}
	\affiliation{Université Grenoble Alpes, CNRS, Grenoble INP, SIMaP
		F-38000 Grenoble, France}
	
	\author{Philippe Jarry}
	\affiliation{C-TEC, Parc Economique Centr'alp, 725 rue Aristide Bergès, CS10027, Voreppe 38341 cedex, France }
	
	\author{Emilie Devijver}
	\affiliation{Univ. Grenoble Alpes, CNRS, Grenoble INP, LIG, F-38000 Grenoble, France}
	
	\author{Thomas Voigtmann}
	\affiliation{Institut für Materialphysik im Weltraum, Deutsches Zentrum für Luft- und Raumfahrt (DLR), 51170 	
		Köln, Germany }
	\affiliation{Department of Physics, Heinrich-Heine-Universität Düsseldorf, Universitätsstraße 1, 40225 Düsseldorf, Germany}
	
	\author{Jürgen Horbach}
	\affiliation{Institut für Theoretische Physik II, Heinrich-Heine-Universit\"at Düsseldorf, Universitätsstraße 1, 40225 Düsseldorf, Germany}
	\author{Andreas Meyer}
	\affiliation{Institut für Materialphysik im Weltraum, Deutsches Zentrum für Luft- und Raumfahrt (DLR), 51170 Köln, Germany }
	\affiliation{Institut Laue-Langevin (ILL), 38042 Grenoble, France}
	\begin{abstract}
		{
In studying solidification process by simulations on the atomic
scale, the modeling of crystal nucleation or amorphisation requires the
construction of interatomic interactions that are able to reproduce
the properties of both the solid and the liquid states. Taking into
account rare nucleation events or structural relaxation under deep
undercooling conditions requires much larger length scales and longer time scales
than those achievable by \textit{ab initio} molecular dynamics
(AIMD). This problem is addressed by means of classical MD simulations
using a well established high dimensional neural network potential trained on a
relevant set of configurations generated by AIMD. Our dataset contains various crystalline
structures and liquid states at different pressures, including their
time fluctuations in a wide range of temperatures considering only their energy labels. Applied to elemental aluminium, the resulting
potential is shown to be efficient to reproduce the basic structural,
dynamics and thermodynamic quantities in the liquid and undercooled
states without the need to include neither explicitly the forces nor all kind of
configurations in the training procedure. The early stage of crystallization is further investigated on a much larger scale with one million atoms, allowing us to unravel features of the homogeneous nucleation mechanisms in the fcc phase at ambient pressure as well as in the bcc phase at high pressure with unprecedented accuracy close to the \textit{ab initio} one. In both case, a single step nucleation process is observed.
		}
	\end{abstract}
	
	\maketitle
	
	\section{Introduction}
	\label{Sec:Introduction}
	
{Apart from steels, aluminium and its alloys
represent the most used and attractive structural metallic materials
due to their specific properties such as low weight, low energy
cost of remelting, and the possibility of almost complete recycling.
Therefore, these materials represent a major axis of the energy
transition \cite{Davis2001}.} An intimate understanding of its
condensed phase properties is founded upon a description of the
atomic level structure and dynamics, and requires an accurate
representation of chemical bonding \cite{Hafner2008}.  This is of
utmost importance in order to tackle phenomena such as phase changes
and solidification process during which a liquid morphs into a solid
either by crystallization or amorphisation \cite{Kelton2010,Royall2015},
showing eventually a change in electronic structure as a
metal-to-semiconductor transition \cite{Jakse2007,Bonati2018}. First-principles approaches, essentially through the Density Functional
Theory (DFT) \cite{Payne1992,Burke2012}, represent the dedicated
framework especially with the breakthrough provided by \textit{ab
initio} molecular dynamics (AIMD) simulations \cite{Car1985} in
combining atomic dynamics with DFT. Despite its enormous success
in many complex chemical bonding environments \cite{Hafner2008},
DFT implementations are limited to a few hundred atoms over time
scales less than $1$ ns \cite{Jakse2013, Pasturel2018} on current
large-scale supercomputing facilities, impeding its use for phenomena
at length and time scales typical of solidification \cite{Sosso2016}.

The desire to bridge typical scales of the electronic structure to
those of the properties under investigation has led to deriving interatomic
potentials with semi-empirical functional forms that average out
or otherwise model electronic degrees of freedom. For metallic
materials of interest here, various approaches were proposed,
starting in the second half of the 20th Century with pair-potentials
based on a nearly-free electron gas description of the electronic
structure \cite{Hafner1987} using simple models within the pseudopotential
theory (PT) \cite{Ashcroft1966,Wils1983,Moriarty1990,Jakse1995}.
It was later acknowledged that it was impossible for pair potentials
to describe on the same footing the structure, dynamic, and
thermodynamic properties in the liquid and solid state
\cite{Belashchenko2013}, with inherent mechanical instability under
shear for crystals. Many-body approaches such as the Embedded-Atom
Model (EAM) \cite{Daw1984,Daw1993}, modified EAM (MEAM) \cite{Baskes1992}
of current widespread use as well as the Reactive Force Field
(ReaxFF) \cite{Huang2019}, just to name a few among many others
\cite{Pettifor1996}, can be considered as successful in this respect.
Fitting the parameters of these potentials is most often oriented
towards describing crystalline phases and transitions between some
of them \cite{Zong2018,Goryaeva2020}, more rarely taking a full
account of the liquid state \cite{Becker2020}. This leads to a lack
of transferability \cite{Belashchenko2013} and a limited ability
to tackle phenomena involving several phases such as crystal
nucleation \cite{Sosso2016}.

{Over the last decade, impressive progress was made
in designing potentials from electronic structure calculations using
supervised Machine Learning (ML) methods
\cite{Behler2015,Behler2016,Ramprasad2017,Schmidt2019,Marques2019,Goryaeva2019,Mueller2020}.
There are now standard libraries for the ML training \cite{Singraber2019}
that can be used in combination with molecular dynamics (MD)
simulation packages \cite{Singraber2019b} such as LAMMPS \cite{LAMMPS}
or in combination with workflow environments such as ASE \cite{Larsen2017}.
On-the-fly ML force field methods have been also proposed \cite{Li2015}
and implemented directly into \textit{ab initio} codes in order to
bypass most of electronic-structure calculation steps \cite{Jinnouchi2019}.
Different ML techniques have been used, ranging from simple linear
regression (LR) methods such as the spectral neighbor analysis
potential (SNAP) method \cite{Thompson2015,Goryaeva2019} to highly
non-linear regression methods using High Dimensional Neural Networks
\cite{Behler2007,Singraber2019} (HDNN) or Kernel Regression (KR)
\cite{Bartok2010,Bartok2015,Botu2017}.} The designed potentials
reach in general an accuracy close to the \textit{ab initio}
calculations from which the database was formed, with exceptional
results for the description of relative stability between crystalline
phases \cite{Morawietz2016} and defects \cite{Goryaeva2020}. However,
approaches taking full account of liquid and
crystalline states remain scarce \cite{Morawietz2016,Bonati2018,Smith2021}
and are often limited to the objective of a good description of the
melting point.  The main reason for this stems from the fact that
the chosen \textit{ab initio} configurations should cover all
situations, as ML techniques may become less reliable outside the
training domain \cite{Behler2015}. It becomes even more crucial for
crystal nucleation occurring under deep undercooling conditions
with a strong evolution of the liquid structure with respect to
that above melting, showing an increasing icosahedral \cite{Jakse2013}
ordering and structural heterogeneity \cite{Pasturel2017} triggering
homogeneous nucleation \cite{Russo2016}.

Machine-learning potentials for aluminium were designed very recently
to describe essentially the properties of the solid states
\cite{Kruglov2017,Bochkarev2019,Smith2021} and the melting temperature
\cite{Kruglov2017,Smith2021}, but none of them taking the liquid structure and dynamics fully into account. Two different approaches were put forward
respectively with a Gaussian kernel regression \cite{Kruglov2017}
and a deep NN \cite{Bochkarev2019} with a dataset built from
configurations extracted from AIMD simulations at various temperatures.
ML potentials were initially trained using the DFT energies starting
from the work of Behler and Parrinello on bulk Si \cite{Behler2007}.
It was subsequently pointed out that the learning process could
benefit from a wealth of additional information if the three
components of the force and six components of the stress per atom
are taken into account \cite{Behler2015,Marques2019}, while one has
only a single energy value per simulated configuration. Still in
some works, only the forces have been used for the training showing
that properties like the vibrational properties in the solid states
can be reproduced, but they remain insufficient to get full account
of thermodynamic quantities \cite{Botu2017,Kruglov2017}. 
Thus whether additional information enhances the training or not can still be questioned, also given the fact that the relative importance of the energy,
forces and stresses for estimating the Mean-Square Error (MSE) or
the Root-MSE (RMSE) introduce two additional free training parameters
\cite{Behler2015}. Moreover, the question of the transferability
of a ML potential taking into account both the liquid and solid
phases as mentioned above, remains essentially unexplored for
aluminium. This aspect is also of importance when dealing with solidification
phenomena.

The aim of the present work is to develop a ML potential for pure
aluminium dedicated to the description of condensed phases, namely
liquid and solid states for temperatures up to $8000$\,K and pressures
up to $300$\,GPa. For this purpose, a HDNN was developed on the
basis of well-known and robust Behler and Parrinello's approach \cite{Behler2007,Behler2015}.
The latter was trained on a data set generated by DFT-based simulations
for the main crystalline structures and liquid states covering the
targeted pressure and temperature domain, including their time
fluctuations by an appropriate sampling of phase space trajectories
\cite{Wales2003}. { It is shown that training the HDNN on sampled AIMD trajectories using solely energy labels leads to an accurate description of the structure, dynamics and thermodynamics  in the investigated domain. More specifically, the single-particle as well as the collective dynamics are well reproduced, which is an essential ingredient in describing solidification. The resulting potential is then applied here to solidification processes, namely amorphisation and early
stages of crystal nucleation, allowing us to unravel features of the homogeneous nucleation mechanisms at ambient as well as high pressure}.

The remaining part of the paper is organized as the following. In
Sec. \ref{Sec:Computational_background}, the specific features of
the HDNN, the training procedure as well as the basic assessment
of the potential on independent DFT and experimental thermodynamic
data are outlined. Sec. \ref{Sec:ResultsDiscussion} is devoted to the test of
the potential's accuracy in describing some structural properties,
the dynamics of the liquid state in the investigated pressure-temperature
domain, as well as the homogeneous nucleation. Finally, in 
Sec.~\ref{Sec:Conclusion}, the main outcomes of the work are given.

\section{Computational background}
\label{Sec:Computational_background}

\subsection{Constructing a machine learning potential}
\label{SubSec:Constructing}
In the last three decades, many potentials of pure
Al and its alloys for the use in atomistic simulations been developed
\cite{Girifalco1959, Jacobsen1987, Mei1992, Ercolessi1994, Mishin1999,
Sturgeon2000, Lee2003, Liu2004, Mendoub2007, Medelev2008, Viney2009,
Zhakhovskii2009, Choudhary2015, Pasquet2015}, using the Morse
potential ansatz \cite{Girifalco1959}, the PT method \cite{Mendoub2007},
embedded-atom method (EAM) \cite{Mei1992, Ercolessi1994, Mishin1999,
Liu2004, Medelev2008, Zhakhovskii2009}, the modified EAM (MEAM)
\cite{Baskes1992, Lee2003, Pasquet2015}, and many-body approaches
such as COMB3 \cite{Choudhary2015}. However, only very few studies
have employed the ML approach \cite{Botu2017, Kruglov2017,
Bochkarev2019}, { none of them taking systematically into account the properties of the liquid state and checking the dynamics which is very important for the solidification aspects. This is precisely one of the aims in building a ML-based potential here.}

Among the various approaches put forward to design ML potentials
\cite{Behler2016,Goryaeva2019}, the choice was made to set up a
high-dimensional neural network built in a similar way to the one
proposed by Behler and Parrinello \cite{Behler2007} and Zhang
\textit{et al.}~\cite{Zhang2018}. {This well established approach
has been proven successful for pure silicon \cite{Bonati2018} as
well as water \cite{Morawietz2016} to model reliably both their
properties in the liquid and solid states.} As a detailed description
can be found for instance in the tutorial review by Behler
\cite{Behler2015} among others, the focus is made mainly on the
specifics of our scheme. The main part consists in a supervised
learning task from a relevant sample of atomic configurations with
known energy generated by AIMD in various crystalline and liquids
structures in the desired temperature and pressure domain. In the
corresponding portion of Potential Energy Landscape (PEL), each
configurational energy is considered as a sum of individual atomic
energies determined from their local atomic environment within a
cut-off radius $r_S$ often extending beyond the first-neighbor
atomic shell, and taken here to be $6.4$\,\AA{} for Al, corresponding to at least the second neighbor shell for all the considered thermodynamic states. This
decomposition allows us to train $N$ Neural Networks (NN), each of
them being assigned to an individual local atomic environment. The
NN are then combined to recover the energy of the whole configuration
of $N$ atoms.

The individual NN is defined by the same network topology for given
atomic species. It specifies the number of neurons formally named
here $y^{l}_{i}$ and their connectivity through the weights,
$w^{l}_{i,j}$, and called a Multi-Layer Perceptron (MLP). The
weights associated with each node pair are optimized during the
learning process by a back-propagation technique \cite{Hastie2009}.
Thus, each of the $M$ layers within the neural network consists of
sets of nodes receiving multiple inputs from the previous layer and
passing outputs to the next layer. Here a fully connected network
is used, in which every output of a layer is an input for every
neuron in the next layer. The corresponding mathematical description
is as follows: the input signals are linearly
combined before being activated by function $f$ to give each output
$y^{l}_i$ of a given fully connected layer $l$ as
\begin{equation}
y_i^{l} = f \left(\sum_{j=1}^{M_{l-1}}w_{i,j}^{l} 
y_j^{l-1}+b_i^{l}\right), 
\label{layer_eq}
\end{equation}
where $M_l$ refers to the size of the $l$-th layer, \textit{i.e.}~the 
number of its neurons. Note that positive weights
enhance connections while negative weights tend to inhibit them.
Most of the activation functions are chosen to have a range in
either $[0,1]$ or $[-1,1]$ and modulate the amplitude of the output.
The activation function $f$ is applied element-wise and is taken
as the softplus function $f(x) = \log(1+e^x)$. Back-propagation
is used to update the network weights and their gradients.

The input layer of a NN takes values representative of one local
atomic environment in the form of a feature vector whose dimension
is then equal to the number of its nodes. The feature vector  is
built on the basis of Behler-Parrinello (BP) descriptors \cite{Behler2007}
to represent the radial and angular arrangements of atoms in the
local structure using Gaussian symmetry functions having the
translational and rotational invariance. For aluminium, the number
of components of the BP feature was chosen to be $22$, comprising
of $12$ radial and $10$ angular components, as described in more
detail in the Supplementary Information (SI) file. Then, the NN
architecture for aluminum is $22\times10\times10\times1$ with $2$ hidden layers of $10$ nodes each.

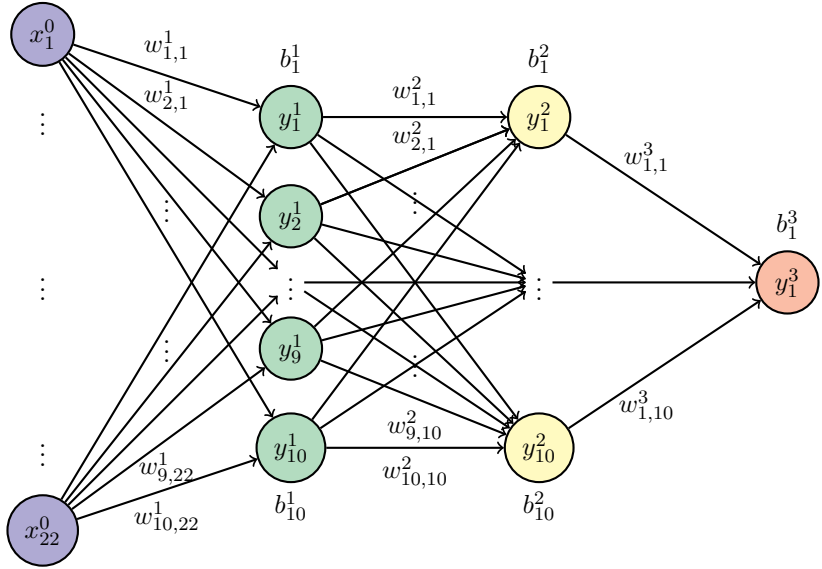
\begin{figure}[t!]
\centering
	
\begin{tikzpicture}[thick,scale=1.1, every node/.style={scale=0.9}]
\node [draw, shape=circle,fill=Blue!30] (x10) at (0,3) {$x_{1}^{0}$};
\node (y30) at (0,2) {$\vdots$};
\node (y40) at (0,0) {$\vdots$};
\node (y50) at (0,-2) {$\vdots$};
\node [draw, shape=circle,fill=Blue!30] (x20) at (0,-3) {$x_{22}^{0}$};

\node [draw, shape=circle,fill=Green!30] (y11) at (3,2) {$y_{1}^{1}$};
\node [draw, shape=circle,fill=Green!30] (y21) at (3,0.8) {$y_{2}^{1}$};
\node (y31) at (3,0) {$\vdots$};
\node [draw, shape=circle,fill=Green!30] (y41) at (3,-0.8) {$y_{9}^{1}$};
\node [draw, shape=circle,fill=Green!30] (y51) at (3,-2) {$y_{10}^{1}$};

\node [draw, shape=circle,fill=Yellow!30] (y12) at (6,2) {$y_{1}^{2}$};
\node (y22) at (6,0) {$\vdots$};
\node [draw, shape=circle,fill=Yellow!30] (y32) at (6,-2) {$y_{10}^{2}$};
	
\node [draw, shape=circle,fill=Red!30] (y13) at (9,0) {$y_{1}^{3}$};

\draw[->] (x10) -- (y11);
\draw[->] (x10) -- (y21);
\draw[->] (x10) -- (y31);
\draw[->] (x10) -- (y41);
\draw[->] (x10) -- (y51);

\draw[->] (x20) -- (y11);
\draw[->] (x20) -- (y21);
\draw[->] (x20) -- (y31);
\draw[->] (x20) -- (y41);
\draw[->] (x20) -- (y51);

\draw[->] (y11) -- (y12);
\draw[->] (y11) -- (y22);
\draw[->] (y11) -- (y32);

\draw[->] (y21) -- (y12);
\draw[->] (y21) -- (y22);
\draw[->] (y21) -- (y32);
	
\draw[->] (y21) -- (y12);
\draw[->] (y31) -- (y22);
\draw[->] (y31) -- (y32);

\draw[->] (y41) -- (y12);
\draw[->] (y41) -- (y22);
\draw[->] (y41) -- (y32);

\draw[->] (y51) -- (y12);
\draw[->] (y51) -- (y22);
\draw[->] (y51) -- (y32);

\draw[->] (y12) -- (y13);
\draw[->] (y22) -- (y13);
\draw[->] (y32) -- (y13);

\node at (1.5,2.85) {$w_{1,1}^{1}$};
\node at (1.5,2.25) {$w_{2,1}^{1}$};
\node at (1.5,0.95) {$\vdots$};
\node at (1.5,-0.75) {$\vdots$};
\node at (1.5,-2.25) {$w_{9,22}^{1}$};
\node at (1.5,-2.85) {$w_{10,22}^{1}$};

\node (b11) at (3,2.7) {$b_1^{1}$};
\node (b51) at (3,-2.7) {$b_{10}^{1}$};

\node at (4.5,2.3) {$w_{1,1}^{2}$};
\node at (4.5,1.75) {$w_{2,1}^{2}$};
\node at (4.5,1.05) {$\vdots$};
\node at (4.5,-0.93) {$\vdots$};
\node at (4.5,-1.75) {$w_{9,10}^{2}$};
\node at (4.5,-2.3) {$w_{10,10}^{2}$};
	
\node (b12) at (6,2.7) {$b_1^{2}$};
\node (b52) at (6,-2.7) {$b_{10}^{2}$};

\node at (7.3,1.5) {$w_{1,1}^{3}$};
\node at (7.3,-1.5) {$w_{1,10}^{3}$};
	
\node (b10) at (9,0.7) {$b_1^{3}$};

\end{tikzpicture}
\caption{Schematic representation of the feed-forward neural network
built as a densely connected multi-layer perceptron. The input
layer $x_i^{0}$ will be fed with the atomic BP feature of the
training set, and the output is the atomic energy. The neural network
contains two hidden layers with superscript $1$ and $2$, respectively.
The two layers are composed of $10$ neurons each. The weights
$w^{k}_{ij}$ and bias $b^{k}_i$ are optimized during the
training (see text).}
\label{fig:FigS1}	
\end{figure}

The NN was coded using \textsc{Keras} module from the \textsc{TensorFlow}
\textsc{Python} package \cite{TensorFlow} in the regression mode.
NNs of all $N$ atoms of a configuration are then associated using
the \textsc{add} module to form the HDNN which is obviously invariant
to permutation of atoms. The HDNN is then trained on the single DFT
energy of the whole configuration.

\subsection{Building the dataset}	
\label{SubSec:Designing}
{Designing an appropriate dataset is the crucial and
demanding step for the construction of the NN interatomic potential. Various strategies can be put forward to construct it, which were reviewed very recently \cite{Unke2021}. 
Here, it was built from AIMD simulations that were partly taken from our
previous works \cite{Jakse2013, Jakse2013a, Jakse2019, Demmel2021}
and extended here to have a better representation of the undercooled
liquid region, the crystalline fcc configurations
up to the melting point at zero pressure, and crystalline fcc, bcc,
and hcp up to $300$\,GPa with and without defects. The different
thermodynamic states and structures as well as the number of
configurations sampled in each case are given in Tables SII and SIII in the
Supplementary Information file. In total, $24300$ configurations
of $N=256$ atoms were gathered in the database that enabled us to
cover solid and liquid states at ambient pressure as well as liquid
samples at temperatures up to 8000\,K and pressures up to 300\,GPa.
Non-equilibrium trajectories in the undercooled region were also
generated to take into account crystal nucleation and solidification
processes in the ML fitting procedure.}

For the sake of self-consistency, the main technical details of the
AIMD simulations are recalled here. They were performed by means
of the Vienna Ab initio Simulation Package (VASP) \cite{Kresse1996}.
The local density approximation (LDA) \cite{Ceperley1980,Perdew1981}
within projected augmented plane-waves was applied to all simulations
with a plane-wave cutoff of $241$\,eV. For the liquid states, only
the $\Gamma$-point is used while for crystalline states, the
$\Gamma$-centered grid of $k$-points in the irreducible part of the
Brillouin zone was set to $2\times2\times2$  following the
Monkhorst–Pack scheme \cite{Monkhorst1976,Blochl1994}. All the
simulations were performed with $N = 256$ atoms placed in a cubic
simulation box (except for the hcp crystal where an orthorhombic box
was used) with standard periodic boundary conditions (PBC). Newton’s
equations of motion were solved numerically with Verlet’s algorithm
in the velocity form with a time step of $1.5$ fs, and phase-space
trajectories were constructed within the canonical ensemble (NVT),
by means of a Nos\'e thermostat to control the temperature $T$. The
temperature evolution in the undercooled states was obtained by
quenching the system stepwise down to $600$\,K with a temperature
step of $50$\,K. For each temperature, the simulation cell was resized
according to the experimental density \cite{Assael2006} and the run
was continued for $30$\,ps before performing the next quench,
resulting in an average cooling rate of $3.3\times10^{12}$\,K/s. The
calculated pressures for all the temperatures studied here were
in the range $\pm1$\,GPa generally, so that on average a quasi constant
pressure during the quenching is observed. For temperatures ranging
from $T = 1000$\,K to $600$\,K, the run was continued for equilibration
during a time up to $200$\,ps. A similar procedure was applied for
heating the fcc crystal from $10$\,K to $900$\,K.

Several aspects deserve attention in the perspective of building
the ML potential. First of all, the choice of the exchange and
correlation (XC) functional for the electronic structure calculations
is crucial. {As the ML potential may reach an accuracy
similar to the DFT calculations}, it will mirror the ability of the
XC functional used in predicting the properties, at least in the
thermodynamic domain inside which it was trained. This has guided
our choice of the Local Density Approximation (LDA) functional given
the fact that the Generalized-Gradient-Approximation (GGA) overestimates
the atomic volume \cite{Alfe2003}. Moreover, it was shown in our
previous contributions that the LDA gives a good description of the
liquid structure \cite{Jakse2013}. More importantly, atomic transport
properties such as the self-diffusion coefficient, which are very
sensitive to the details of the potentials, are well reproduced
within the LDA compared to state-of-the-art experimental data
\cite{Demmel2011,Kargl2012}. Such a good agreement with experiments
was very recently confirmed on the dynamic structure factors as
well as the structural relaxation times extracted from the
intermediate scattering function \cite{Demmel2021}. For high
pressures, it was shown \cite{Sjostrom2016} that the difference
between LDA and GGA \cite{Perdew1992} is negligible in describing
the pressure-density phase diagram of aluminium up to pressures as
high as $10$\,TPa.

{Secondly, in the perspective of performing MD
simulations, care has to be taken in describing not only average
thermodynamic properties but also the fluctuations around the mean
value{, especially in order to capture the features of local basins
of the PEL \cite{Wales2003}.}  This requires the sampling of a large
number of configurations along AIMD phase space trajectories.
Therefore, in the present work, for each of the considered thermodynamic
states (see Tables SII and SIII of the SI file), $1000$ configurations were
generated on AIMD production runs over $40$\,ps.
}

Finally, as mentioned in the introduction, the question whether
including the additional information of the forces or even the
stresses in addition to the energies improves the learning process
and the accuracy of the potential deserves further attention. It was shown very recently for molecular systems that forces and energies contribute equally to the convergence of the prediction errors \cite{Christensen2020}. { The choice of considering energies, forces or both of them in the training  may depend on factors such as the application domain, the properties of interest, the complexity of the ML tool, and the strategy in building the data from \textit{ab initio} calculations. When making static DFT calculations on chosen configurations, including forces and/or stresses labels make more sense, especially when augmentation of information is performed by generating configurations from it by random atomic displacements. Here a strategy solely based on energy labels for the training is chosen since the data consists of sampled AIMD trajectories for each thermodynamic state whose accessible microstates explore, through the thermal fluctuations, their local basin of the PEL, taking implicitly their gradients into account \cite{Wales2003}.}

\subsection{Training the Neural Network}
\label{SubSec:Training}
{The supervised training is carried out using as
input the BP feature vectors describing local atomic environments
in each configuration. AIMD energies of these configurations are
used to find the optimal set of weights and biases. The complete
dataset of configurations is firstly randomized and scaled using
the standard scaler of \textsc{Scikit-learn}, \textit{i.e.}~centering
the feature components about their mean and normalize them according
to their standard deviation. It is then split into a training set
of $80$\% of the data and a test set containing the $20\%$ remaining
part. In the training set $20$\% of the data are retained further to create
validation sets. They are used (i) for a cross-validation procedure
to estimate the performance of various NN architectures through the
MSE, and (ii) to monitor the MSE on the validation data during the
learning process to detect overfitting. For a given architecture,
the optimization is performed using the training data without the
validation set, and terminating when the validation error starts
to increase. Reduction of the noise of the MSE during training is
obtained by including a callback with a stepwise reduction of the
learning rate. Simultaneously, a $L_2$ norm regularization with strength $10^{-5}$ is
performed to reduce the model complexity, and thus to prevent
overfitting. Once trained, the weights and biases are stored in a
format compatible with the \textsc{LAMMPS} HDNNP pair-style
\cite{Singraber2019b}.}

\begin{figure}[t!]
\centering
\includegraphics[scale=0.46]{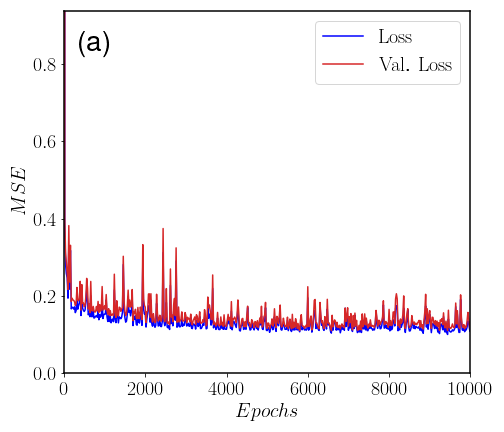}
\includegraphics[scale=0.46]{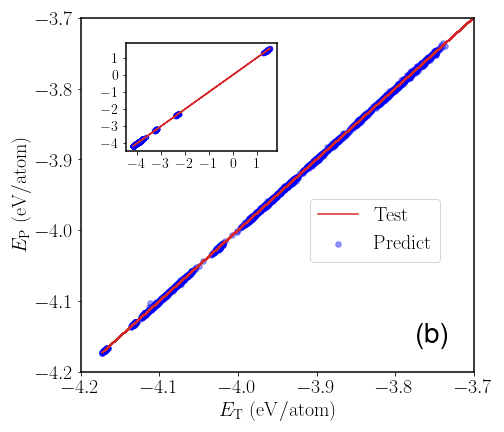}
\caption{(a) Evolution of the MSE losses as a function of number
of epochs for the training and validation sets. Loss and Val. Loss correspond to the evaluation of the MSE on the training set and validation set, respectively. (b) Test-Predict
curve showing the quality of the prediction on the test set for the
optimized architecture $22\times10\times10\times1$. The red solid
line represents the known output of the energies representing the know $E_T$=$E_T$ line that would correspond to a perfect prediction, and the blue dots
the values predicted against the known ones. The main panel corresponds
to thermodynamic states at low pressures ($<5$ GPa) and temperatures
between $10$\,K and $1500$\,K containing either crystalline and liquid
configurations. The inset corresponds to energies in the full range
of temperatures and pressures (see text).}
\label{fig:Fig1}
\end{figure}

This training stage is repeated with various NN architectures to
find the optimal one capturing at best the functional dependence
of the data. Evaluation of the MSE is carried out through a stochastic
gradient descent minimization using the Adam optimization algorithm
\cite{Hastie2009} giving a measure of the loss with a learning rate
starting at $0.01$ and reducing most of the time to $0.0001$ during
the training, $\beta_1=0.9$, $\beta_2=0.999$ and $\varepsilon=10^{-8}$.
The early-stopping was performed with maximum loss variation of
$10^{-6}$ and a patience of $45$ epochs. The typical duration of
the training period was about $10000$ to $15000$ epochs. The least
MSE loss is obtained for an architecture of $10$ neurons in the
first and second hidden layers. A typical evolution
of loss and the validation loss is shown in Fig.~\ref{fig:Fig1}(a).
A cross-validation performed over $5$ independent trainings gives
a RMSE of $(1.2)$\,meV on the per atom energy. Figure \ref{fig:Fig1}(b)
displays the predictive ability of the model on the unseen data of
the test set, with a high quality over the whole range of the
energies.

\begin{figure}[t!]
\centering
\includegraphics[scale=0.29]{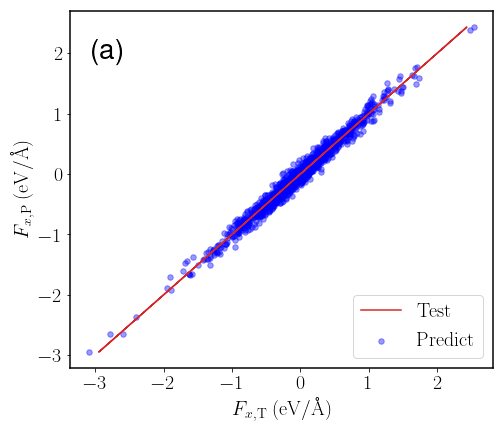}
\includegraphics[scale=0.29]{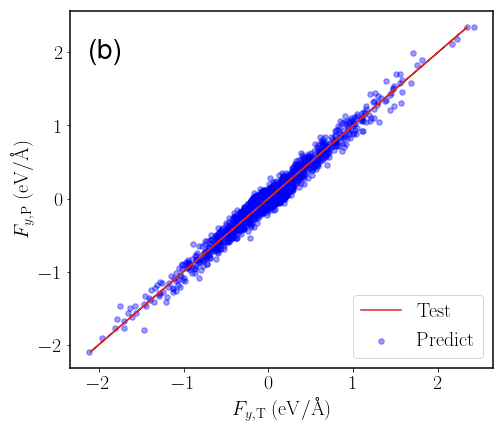}
\includegraphics[scale=0.29]{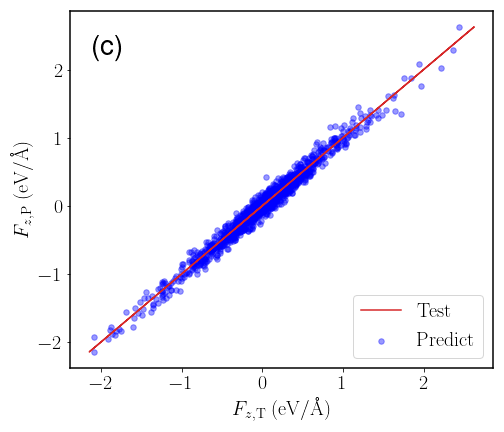}
\caption{Test-Predict curves for the three components of the predicted
forces against those extracted from the AIMD configurations at
$T=1500$\,K over $2$\,ps. The red solid line and blue dots have the same meaning as in Fig. \ref{fig:Fig1}.}
\label{fig:Fig2}
\end{figure}

The predictive ability of the HDNN is illustrated in Fig.~\ref{fig:Fig2}
on the three forces components extracted from AIMD configurations
of a simulation at $1500$\,K over $2$\,ps, with a RMSE of $0.074$\,eV/\AA{} at this high temperature. At $T=10$\,K in similar sampling conditions the RMSE reduces to $0.030$\,eV/\AA{}. These RMSE values are consistent with those obtained for previous trained ML potentials for Al for which forces were included explicitly in the training \cite{Kruglov2017,Bochkarev2019}. Our results lead to similar conclusions for molecular systems \cite{Christensen2020} saying that the forces    
can be predicted with a good accuracy without being explicitly part
of the learning process, and thus is in favor of the supervised
learning strategy based only on the energy of the configurations, thus avoiding additional parameters in the loss function.

\begin{figure}[t!]
\centering
\includegraphics[scale=0.70]{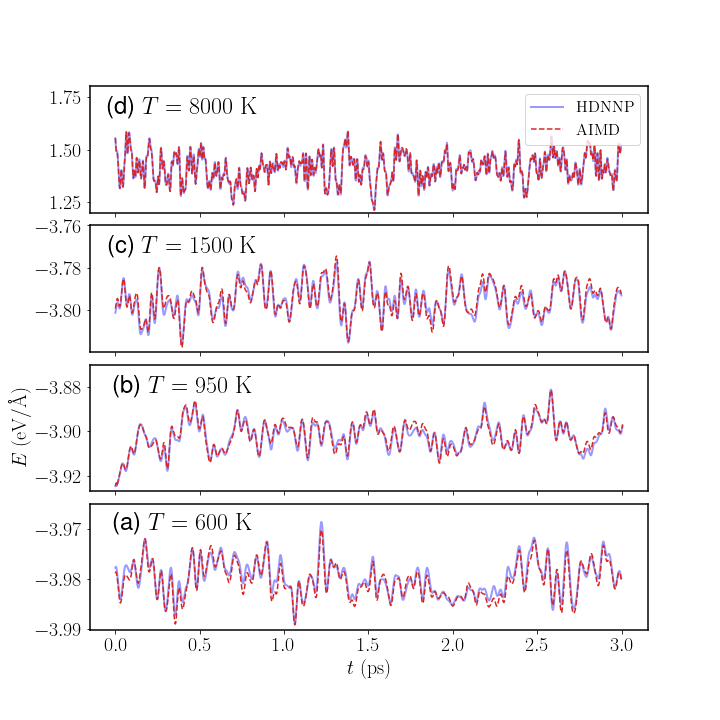}
\caption{Energy per atom as a function of time from AIMD simulation
of liquids and predicted by the HDNN potential at $T=600$\,K in the
undercooled region, $T=950$\,K in the vicinity of the melting point,
at $1500$\,K far above the melting point at zero pressure, and
$T=8000$\,K just above the melting line for a pressure of $322$\,GPa.}
\label{fig:Fig3}
\end{figure}

The HDNN is further tested on the prediction of the energy as a
function of time. Consecutive configurations of AIMD simulation of
liquids were considered at $T=600$\,K in the undercooled region,
$T=950$\,K in the vicinity of the melting point, at $1500$\,K far
above the melting point at zero pressure, and $T=8000$\,K just above
the melting line for a pressure of $322$\,GPa, as show in 
Fig.~\ref{fig:Fig3}. Energy fluctuations are very well reproduced for all 
the temperatures, even for the extreme values for which the sampling
is scarce, as they correspond to the tail of the Gaussian distribution
of energy fluctuations. Probably the most impressive agreement is
that of the simulation at $T=8000$\,K where the energy range of the
fluctuation is as large as $0.25$\,eV/atom and still very well
predicted. This demonstrates the quality of the ML potential.

\subsection{Molecular dynamics simulation}
\label{SubSec:MDsimulations}
%
Classical MD simulations were carried out using the
\textsc{LAMMPS} package \cite{LAMMPS}. These simulations were
performed in various ensembles, namely the canonical ensemble
($NVT$, constant temperature, volume, and number of atoms), the
isobaric-isothermal ensemble ($NPT$, constant temperature, pressure,
and number of atoms), and the isobaric-isoenthalpic ensemble (NPH,
constant pressure, enthalpy, and number of atoms). Temperature and
pressure were kept constant \textit{via} the Nose-Hoover thermostat and
barostat \cite{All1989,Smi2002}, respectively. In all simulations,
PBC were employed in the three spatial directions. The integration
of the equations of motion was done \textit{via} Verlet's algorithm in the
velocity form, choosing a time step of $1$\,fs. The use of our HDNN
potential in \textsc{LAMMPS} was possible through the library-based
implementation of high-dimensional neural network potentials by
Singraber \textit{et al.}~\cite{Singraber2019b}. To assess the quality of our HDNN potential, some of the MD simulations were also repeated with the previously published ANI-Al potential of Smith \textit{et~al.}~\cite{Smith2021}.

Structural analysis is performed using the common-neighbor
analysis (CNA) \cite{Hon1987} with the indexing of Faken and Jonsson
\cite{Faken1994} and a bond-based algorithm as implemented in the
OVITO software \cite{Stu2010} where a uniform cut-off radius
corresponding to the first minimum of the pair-correlation function
of the liquid is applied to create bonds between pairs of particles.
The CNA classifies pairs around each atom by sets of three indices:
the first index represents the number of nearest-neighbors common
to this pair, the second index corresponds to the number of
nearest-neighbor bonds among the shared neighbors, and the third
index indicates the longest chain of bonded atoms among them. For
instance, $421$ and $422$ bonded pairs are characteristic of close
packed structures fcc and hcp, respectively. The occurrence of $444$
and $666$ pairs, with specific proportions, signals the presence
of bcc ordering. The degree of five-fold symmetry is obtained from
the proportion of $555$, $554$ and $433$ pairs, which represent
perfect ($555$) and distorted FFS based motifs.

An alternate way of studying the local ordering before and during nucleation is to make use of the Steinhardt bond-ordering parameters\cite{Steinhardt1983}. More specifically,  the averaged form\cite{Dellago2008} as implemented in the \textsc{Pyscal} code \cite{pyscal2019} is considered here. First, for each atom $i$,  the following vector is define 
\begin{equation}
q_{lm}(i) = \frac{1}{N(i)} \sum_{j=1}^{N(i)} Y_{lm}(\mathbf{r}_{ij})
\end{equation}
where $N(i)$ is the number of nearest neighbors of atom $i$, $\mathbf{r}_{ij}$ is the displacement of nearest-neighbor atom $j$ from $i$, and $Y_{lm}$ is the spherical harmonics. From these, the averaged bond-order parameters can be defined as
\begin{equation}
\bar{q}_{l}(i) = \sqrt{\frac{4\pi}{2l + 1}\sum_{m=-l}^{l}
	\left|\frac{1}{N(i)+1}\sum_{k=0}^{N(i)}q_{lm}(k)\right|^2}
\end{equation}
where the sum from $k=0$ to $N(i)$ includes both the atom $i$ and its nearest neighbors.
Due to being averaged over nearest neighbors, these parameters take into account not just the first coordination shell, but also the second one. To perform structural analysis using these parameters one typically selects specific values for $l$, with $l=4$ and $l=6$ being a common choice.
It is then possible to compare the resulting values of $\bar{q}_l$ to those of ideal crystals in order to identify crystal structure.

\section{Results and discussion}
\label{Sec:ResultsDiscussion}
\subsection{Local structure and dynamics}
\label{SubSec:Dynamics}
In a first step, the optimized HDNN potential is assessed on the
local structure and dynamics.
The simulations are performed at a constant volume, given in top row of Table \ref{tab:ndp}, with $N\simeq 10000$ atoms
at selected temperatures.
For the undercooled states the system is first prepared at a temperature of $1500$\,K, before being cooled down to the desired temperature.
The other states are prepared directly at the target temperature from a fully equilibrated liquid.
Following an equilibration time of $10$\,ps relevant quantities are calculated
over a production time ranging from $100$\,ps to $1$\,ns depending on the thermodynamic
state under consideration.

\begin{figure}[t]
\centering
\includegraphics[scale=0.215]{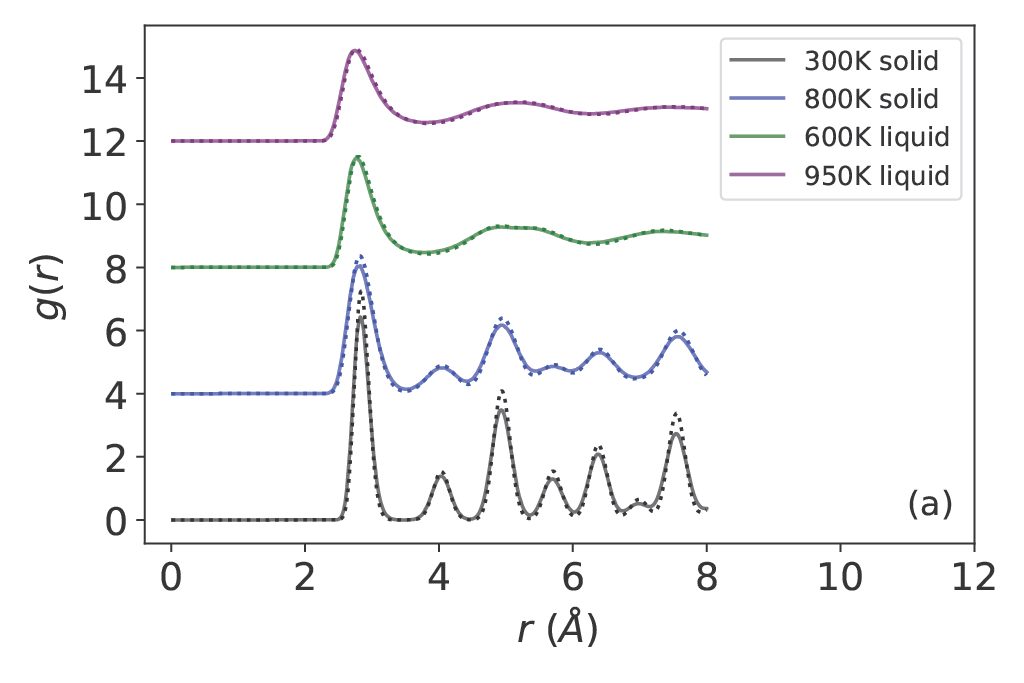}
\includegraphics[scale=0.21]{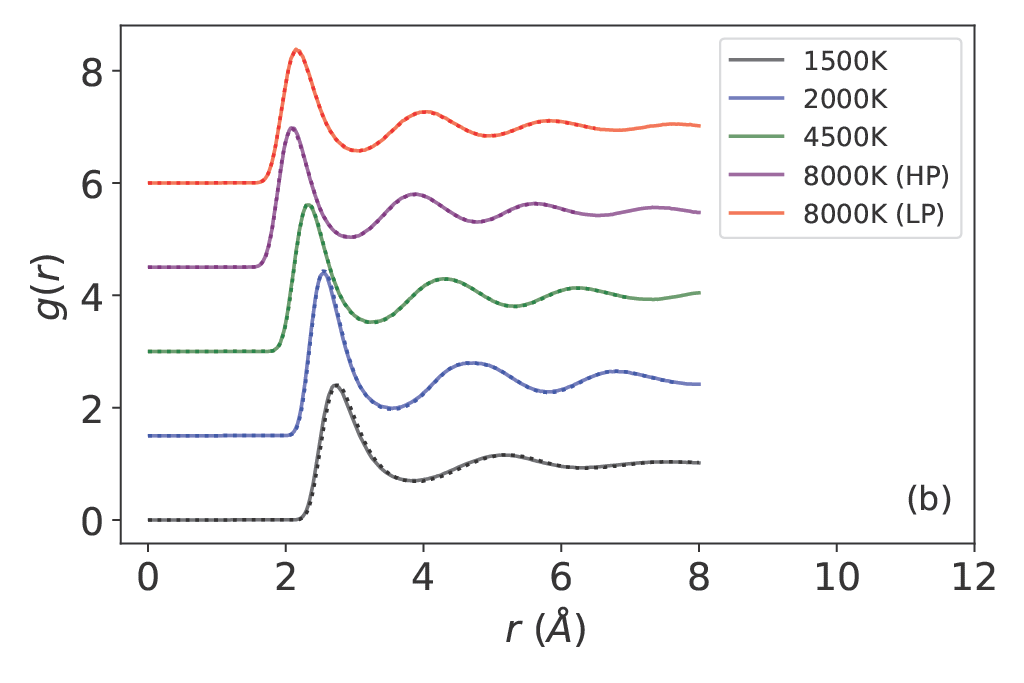}
\includegraphics[scale=0.21]{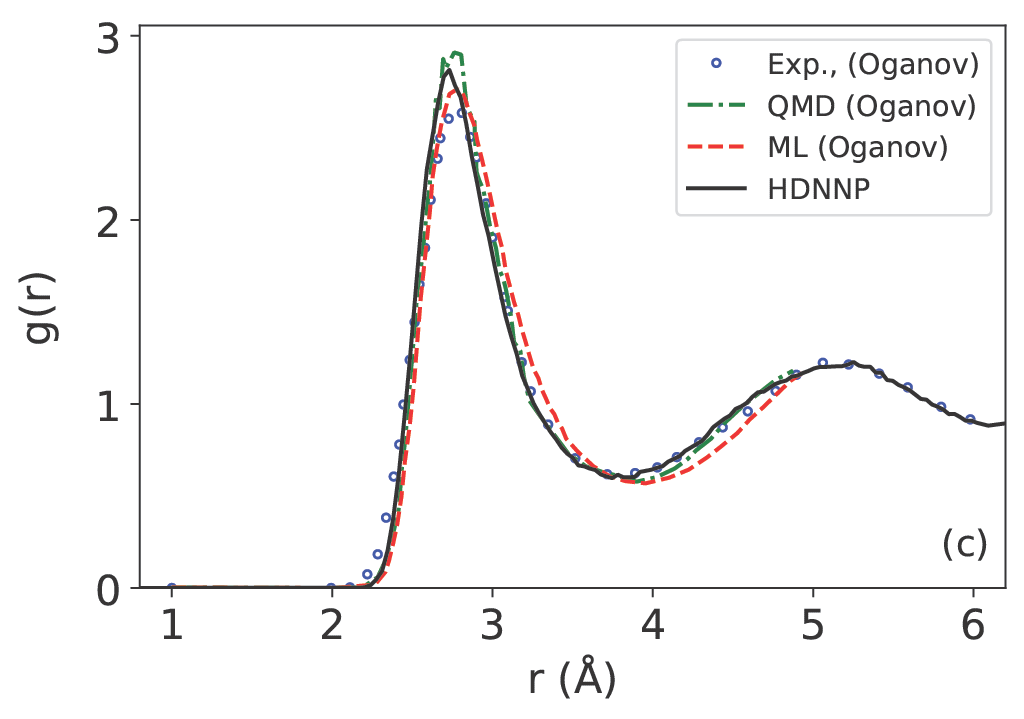}
\includegraphics[scale=0.204]{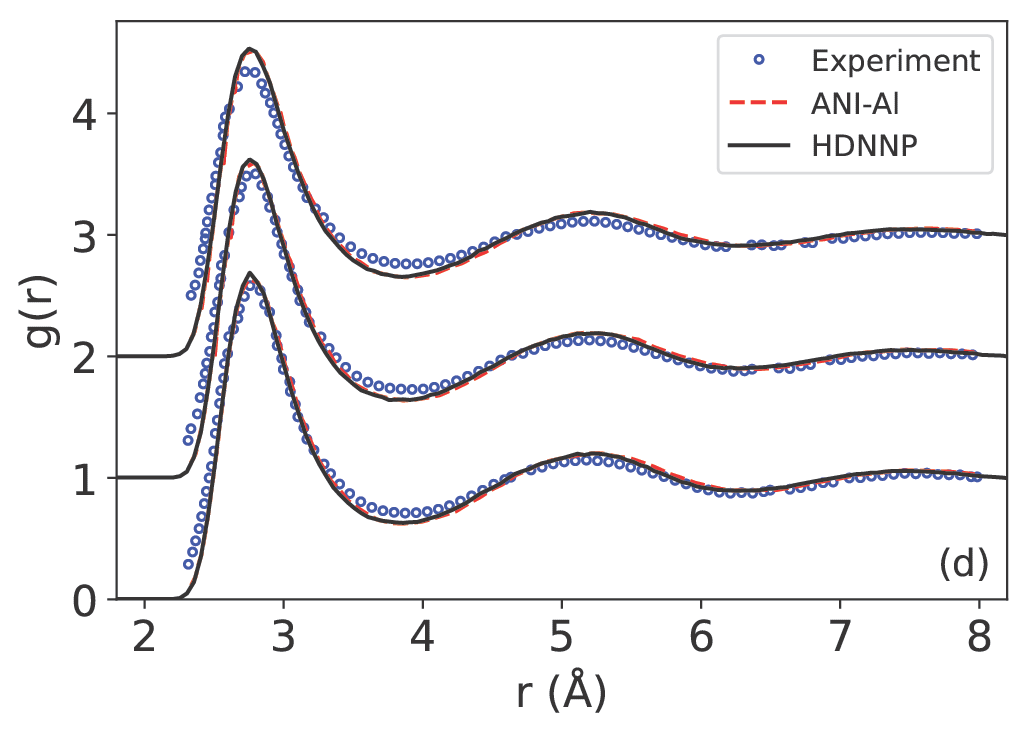}
\caption{\label{fig:g}
Pair-correlation function for various temperatures and pressures
(a) for low temperature liquid and solid states at room pressure
and (b) for high temperature and pressure liquid states. Curves for
$800$~K, $600$~K, $950$~K, $2000$~K, $4500$~K, $8000$~K ($340$~GPa)
and $8000$~K ($240$~GPa) are shifted upwards by an amount of $4$,
$8$, $12$, $1.5$, $3$, $4.5$ and $6$, respectively. The solid lines
are results with the HDNN potential, the dotted lines to the ANI-Al potential, and the dashed lines with
corresponding colors are those of the AIMD simulations. (c) Comparison  in the liquid at $T = 1023$ K to simulations with the ML potential of Ref. \cite{Kruglov2017} as well as their \textit{ab initio} calculations and experimental data. (d) Comparison in the liquid at $T=1123$~K, $T=1183$~K and $T=1273$~K to simulations with the ANI-Al potential of Ref. \cite{Smith2021} as well as experimental data of Ref. \cite{Mauro2011}.}
\end{figure}

The pair-correlation function $g(r)$ gives the probability of finding a particle $j$ at
distances $r_{ij}$ relative to a particle $i$ located at the origin,
and reads:
\begin{equation}
g(r_{ij}) = \frac{N}{V}\frac{n(r_{ij})}{4\pi r_{ij}^{2}\Delta r}.
\label {Eqpcf}
\end{equation}
$n(r)$ represents the mean number of particles $j$ in a spherical
shell of radius $r$ and thickness $\Delta r$ centered on particle
$i$.
Finally, an average of $g(r)$ over all $N$ particle $i$ of the
simulation box is performed. Integrating $4\pi r^2\rho g(r)$, with
$\rho=N/V$ up to the first minimum $g(r)$ gives access to the mean
coordination number.
Figure \ref{fig:g} displays the curves of $g(r)$ from $NVT$ simulations. An excellent match with AIMD simulations is seen for all the thermodynamic states.
A quantitative estimation of the deviation was obtained by calculating the MSE
between the classical MD and AIMD curves for each temperature.   
The MSE ranges from $6\times10^{-4}$ typically in the case of the liquid states to
$1.2\times10^{-2}$ in the case of fcc solid states. The larger MSE
for the solid might come from the fact that even a very small
position shift of $g(r)$ can induce a significant deviation as peaks
are sharp and narrow. { The curves of $g(r)$ obtained from the ANI-Al ML potential of Smith \textit{et al.} \cite{Smith2021} are slightly shifted to larger distances for all liquid states considered, thus overestimating the bond lengths, and their peaks in the crystalline states are more pronounced. The average coordination numbers with the HDNN potential display} a
deviation from AIMD that does not exceed $0.2$ (see Table \ref{tab:ndp}).
{ Nevertheless, a comparison of $g(r)$ of the
present HDNN potential to those obtained with published ML potentials is considered.
Fig. \ref{fig:g}(c)  shows that our potential leads to overall better results than
the ML potential of Kruglov \textit{et al.} \cite{Kruglov2017} as compared to
their \textit{ab initio} simulations and experimental data at $T=1000$~K.
In Fig. \ref{fig:g}(d) our potential leads to results very close to the recent
ANI-Al potential of Ref.~\cite{Smith2021}, and in good agreement
with experiments \cite{Mauro2011} at $T=1123$~K, $T=1183$~K and $T=1273$~K.
It is worth mentioning that both mentioned ML potentials have been trained
using only the forces \cite{Kruglov2017} or using the energies and forces \cite{Smith2021},
contrary to the present potential.
Additional }comparison with the most widely used EAM \cite{Medelev2008}
and MEAM \cite{Lee2003} potentials is shown in Fig.~S1 in the Supplementary
Information File for the same thermodynamic states.
They are shown to perform less well than the HDNN potential,
as assessed by a $t$-statistics, and especially for the high pressures.
	
\begin{table}[t]
\begin{center}
\begin{tabular}{cccccccccc}
	\hline\hline
	$T (K)$ & $300$ (s) & $600$ & $800$ (s) & $950$& $1500$& $2000$& $4500$& $8000$ (H) & $8000$ (L)\\
	\hline
        $V$ (\r{A}$^3$) & $16.48$ & $18.06$ & $16.68$ & $18.89$ & $20.37$ & $14.00$ & $10.67$ & $7.629$ & $8.616$\\
        $P$ (GPa) & $0$ & $0$ & $0$ & $0$ &$0$ &$30$ &$110$ & $240$ & $340$ \\
	& $0$ & $0$ & $0$ & $0$ & $0$ & $27$ & $107$ & $227$ & $320$\\
	$N_C $ & - & $11.93$  & - & $11.57$  & $11.10$  & $12.22$ & $12.70$ & $12.70$  & $12.65$ \\
	& - & $12.05$ & - & $11.65$ & $11.17$ &  $12.32$ & $12.80$&  $12.80$ & $12.75$\\
	$D$ (\AA{}$^2$/ps) & - & $0.15$  & - & $0.63$ & $1.69$  & $0.35$ & $0.78$ & $1.01$  & $1.35$ \\	
		& - & $0.14$ & - & $0.62$ & $1.61$ & $0.37$ & $0.68$ & $0.93$ & $1.34$ \\
		& - & ($0.10$) & - & ($0.46$) & ($1.54$) & ($0.33$) & ($0.66$) & ($0.96$) & ($1.38$) \\
	\hline\hline				
\end{tabular}
\end{center}
\caption{\label{tab:ndp}
Atomic volume $V$, pressure $P$, coordination number $N_C$, and diffusion coefficient $D$
for selected temperatures. Values in second rows are from the AIMD, and those in parenthesis are the diffusion coefficient calculated from the ANI-Al potential \cite{Smith2021}. Classical MD simulations for the diffusion were performed with $N=256$ atoms as for the AIMD one in order to have similar finite size effects. (H), (L), and (s) specifies high and low pressure, and solid state simulations respectively.}
\end{table}
	
\begin{figure}[t]
\centering
\includegraphics[scale=0.205]{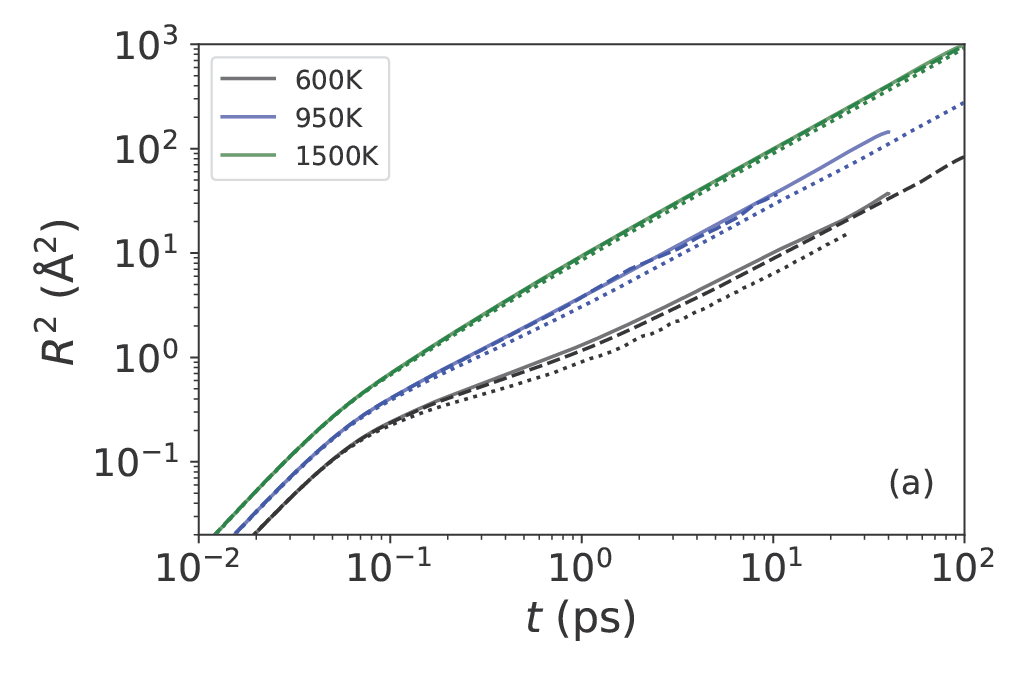}
\includegraphics[scale=0.205]{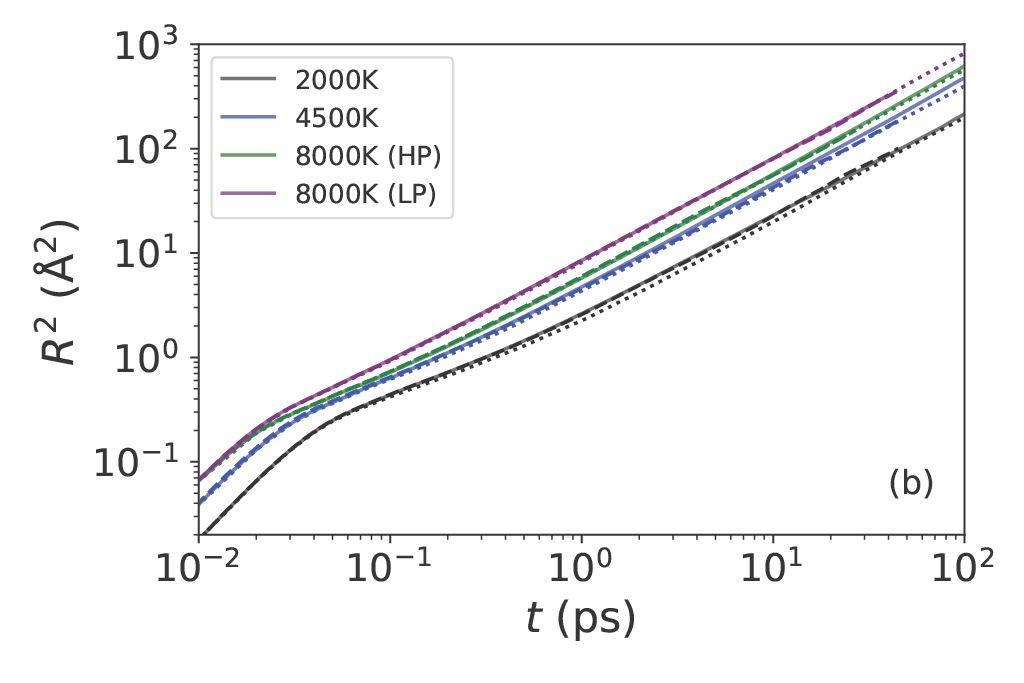}
\includegraphics[scale=0.36]{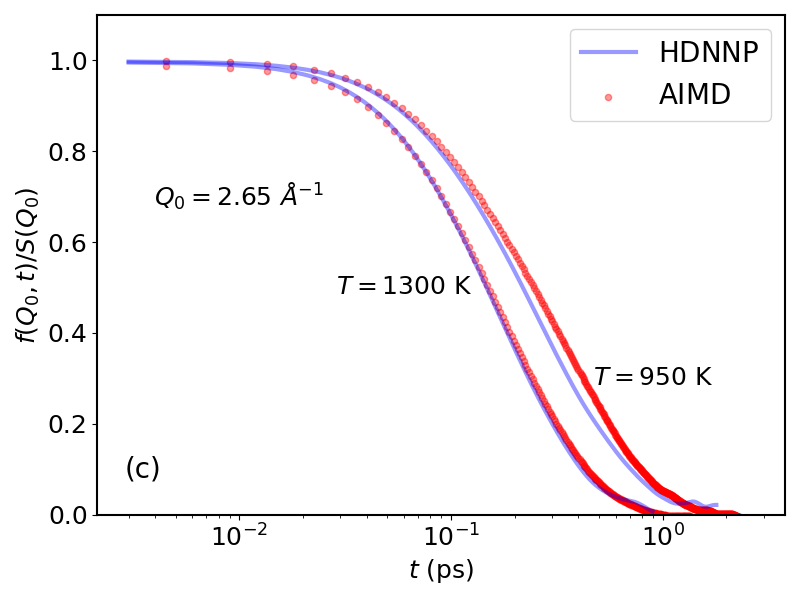}
\includegraphics[scale=0.37]{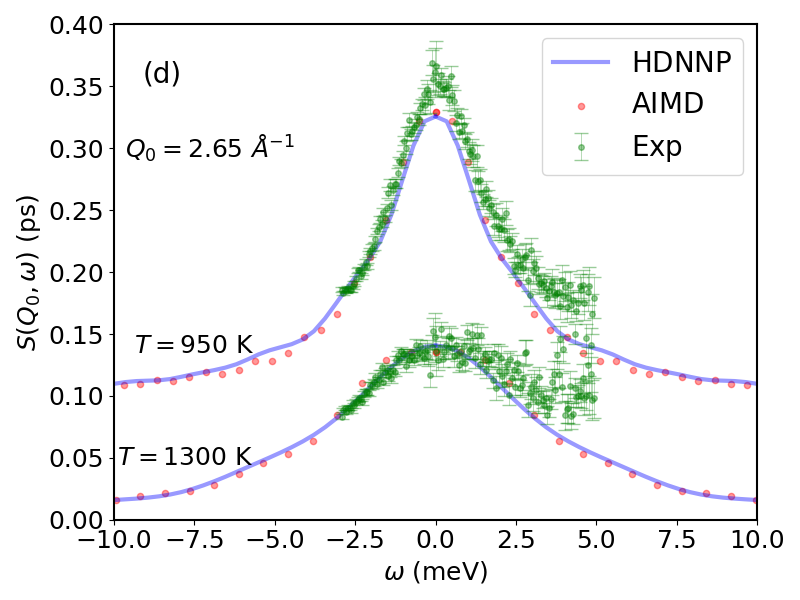}

\caption{\label{fig:r}
Mean-square displacement for various temperatures and pressures: (a)
for high temperature and pressure liquid states and (b) for low
temperature liquid and solid states at room pressure. The solid
lines correspond to the HDNN potential, the dotted lines to the ANI-Al potential~\cite{Smith2021}, and the dashed lines are the AIMD simulations. Classical MD simulations for the MSD were performed with $N=256$ atoms as for the AIMD one in order to have similar finite size effects. (c) Intermediate scattering function in the liquid state at at $T = 950$~K and $T = 1300$~K for wave vector $Q_0=2.65$ \AA{}$^{-1}$. (d) Corresponding dynamic structure factor at the same temperatures and wavevector that are compared to the neutron diffraction experiments at $T = 943$~K and $T = 1293$~K from Ref. \cite{Demmel2021}.}
\end{figure}

Beside the local structural properties, dynamic properties represent
a stringent test as they are even more sensitive to the details of
the potentials. Among these, diffusion plays an important role in
the solidification process \cite{Sosso2016,Kob2005} and was evaluated
here through the mean-square displacement (MSD)
\begin{equation}
R^{2}(t)=\frac{1}{N}\sum_{l=1}^{N}\left\langle \left[ \mathbf{r}%
_{l}(t+t_{0})-\mathbf{r}_{l}(t_{0})\right] ^{2}\right\rangle _{t_{0}},
\label {EqR}
\end{equation}
where {$\mathbf{r}_{l}(t)$} denotes the position of atom $l$ at
time $t$ and $N$ is the number of atoms. In addition to the mean
over all atoms, an averaging over time origins $t_{0}$ as indicated
by the angular brackets is performed. The self-diffusion coefficient
$D$ is determined from the slope of the linear behavior at long
times of the MSD. In Fig.~\ref{fig:r}, the MSD is shown for
temperatures in the stable and undercooled liquid states at ambient
pressure as well as for temperatures along the melting line for
pressures up to $340$\,GPa. The overall trends of AIMD curves are
well reproduced by the HDNN potential for all temperatures and
pressures as can be seen in Table \ref{tab:ndp}. { It is worth mentioning that our potential gives a better prediction of the diffusion coefficients than the ANI-Al potential at ambient pressure and low temperatures where solidification phenomena occur.}  The MSD curves show a ballistic regime at very short times ($t<0.05$\,ps), followed by a diffusive regime at long times. For the
lower temperatures at ambient pressures and at high pressures a
well-known caging effect \cite{Kob2005} takes place after the
ballistic motion and delays the diffusive regime, which is well predicted
by the HDNN potential with respect to the AIMD. 

{ The collective dynamics is examined by means of the Intermediate Scattering Function (ISF) $F(Q,t)$ and its time Fourier transform $S(Q,\omega)$, the dynamic structure factor that can be measured by means of Neutron Diffraction. $Q$ represents the wave-number and $\omega$ the frequency. Fig. \ref{fig:r}(c) shows the good agreement of the ISF between the HDNN potential and the AIMD results in the liquid state at $T = 950$~K and $T = 1300$~K for $Q_0=2.65$ \AA{}$^{-1}$ corresponding to the position of the first maximum of the static structure factor $S(Q)$. The good match is confirmed for $S(Q,\omega)$ in Fig. \ref{fig:r}(d) for both temperatures. Further, the comparison with Neutron diffraction data \cite{Demmel2021} demonstrates that the \textit{ab initio} calculations as well as the HDNN potential predict the dynamic properties of liquid aluminium quite accurately.}

The real predictive character of the HDNN potential is assessed for
$T = 950$~K,  $T = 2000$~K and $T = 8000$~K at the lowest pressure:
These thermodynamic states were not included in the training,
but the accuracy is similar to the other states for the pair-correlation
function and the diffusion. { This is all the more true for results at $T=1123$~K, $T=1183$~K, $T=1273$~K and $T = 1300$~K for the comparison with other ML potentials and experiments in Figs. \ref{fig:g} and \ref{fig:r}, as well as for }the pressure for all states shown in Table \ref{tab:ndp} with a deviation 
less than $7$\% at the highest ones, which is remarkable since the forces were not included in the training.

\subsection{Thermodynamic properties}
\label{SubSec:GlassTransition}
\begin{figure}[t]
\centering
\includegraphics[scale=0.225]{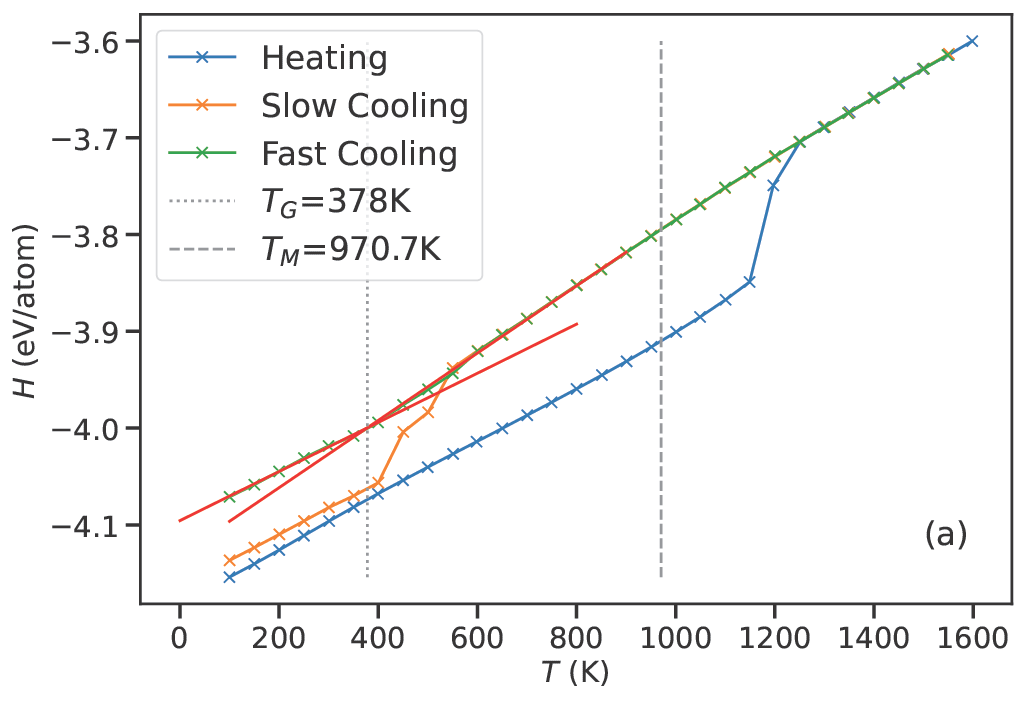}
\includegraphics[scale=0.225]{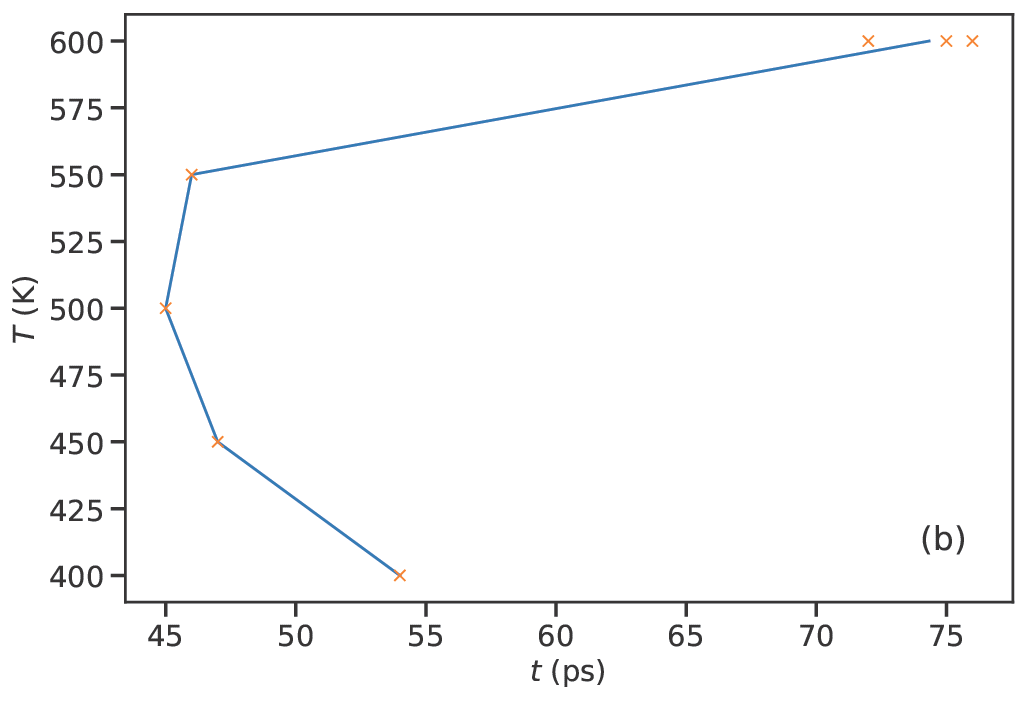}
\caption{\label{fig:H}
(a) Enthalpy as a function of temperature for the solid (from heating)
and liquid branches (from cooling) at ambient pressure as described
in the text. The red lines indicate the slope of the amorphous and
the supercooled states as a guide for the eyes for the crossover
between the liquid and amorphous regimes, and marked by the vertical
dotted line. The vertical dashed line marks the melting temperature
obtained from the the LSI simulations.
(b) The TTT curve, as described in the text. The orange crosses are individual measurements of the time it takes to reach $40\%$ solidification, with the blue line connecting the averages for each temperature.}
\end{figure}

Important quantities for solidification phenomena are the latent
heat of fusion \cite{Herlach2015,Orava2014} as well as the densities
of the solid and liquid phases at the melting temperature, $T_M$.
Its determination requires the calculation of the enthalpy difference
between the liquid and solid branches at $T_M$. The temperature evolution of the enthalpy at ambient pressure
for the solid and liquid branches of the HDNN potential are shown
in Fig.~\ref{fig:H}(a), obtained \textit{via} simulation of $N\simeq 2000$ atoms in the
NPH ensemble.
The simulation is started with a perfect fcc
crystal at $T=300$~K and heated stepwise with a temperature step
of $50$~K with an average heating rate of $10^{12}$~K/s. At each
temperature, a simulation is performed over $50$\,ps ($25$\,ps
equilibration and $25$\,ps production) during which an average value
of the enthalpy is calculated. The increase of temperature is
repeated until a dynamic melting is observed at
$T=1250$~K. The latter value is noticeably higher than the thermodynamic
melting temperature $T_M=970$~K obtained from LSI simulations due
to overheating effects. For the liquid branch, the simulations are
started at $T=1600$~K with an equilibrated configuration after the
heating  process. The same procedure as for the solid branch is
followed but with a step-wise cooling down to $300$\,K that is called here
the slow cooling. Above $T=1250$~K, the difference in the enthalpy
from the heating and cooling processes is negligible, indicating
that there is no reminiscence of the crystalline state. Below $600$\,K 
the liquid undergoes a partial crystallization during cooling.
Then, the cooling procedure for the liquid branch is repeated with
a higher cooling rate of $10^{13}$~K/s to avoid crystallization,
and a glass transition is seen at $T_G=378$~K inferred from a
crossover between the liquid and glassy branches as shown in 
Fig.~\ref{fig:H}(a). 

For such a high cooling rate of $10^{13}$~K/s, the time to reach each temperature  is shorter than its corresponding nucleation time. This is illustrated from the time temperature transition (TTT) curve plotted in Fig. \ref{fig:H}(b). It is determined for a system of approximately $1$ million atoms (see Sec. \ref{SubSec:Nucleation})  by measuring the time it takes until $40\%$ of the atoms are identified as in a solid state by CNA. At $600$~K, the measurement is repeated from the initial configuration, but with new velocities picked from a proper Maxwell distribution, to estimate the variability between measurements.

From the liquid and solid branches an enthalpy of melting of $11.67$~kJ/mol is found, which compares reasonably well to
the experimental value of $13.34$~kJ/mol \cite{Leitner2017}. Taking
the numerical derivative of the solid branch yields a value of the
specific heat at constant pressure, $C_P$, of $0.99$~J/g/K which
is also in good agreement with the experimental value of $0.91$~J/g/K.
For the liquid a value of $1.158$~J/g/K is obtained, which is in
the range of experimental data between $1.03$~J/g/K and $1.18$~J/g/K
close to the latest assessed values of $1.127$~J/g/K \cite{Leitner2017}.
The specific heat is a typical derivative quantity that depends on
the fluctuations of the enthalpy \cite{All1989,Smi2002}. The very
good agreement is a strong indication that including the time
fluctuations from AIMD is a fruitful strategy to describe at least
the thermodynamics.

Regarding densities, the HDNN potential gives a value of 
$0.0587$\,\AA{}$^{-3}$ and $0.0547$\,\AA{}$^{-3}$, respectively in the solid
and liquid at its melting point $T_M=970$~K, giving rise to a density
change of $0.004$\,\AA{}$^{-3}$. These values compare well to the
respective experimental values \cite{Simmons2009,Assael2006,Leitner2017}
of $0.0573$\,\AA{}$^{-3}$ and $0.05306$\,\AA{}$^{-3}$ with a density
change of $0.0042$\,\AA{}$^{-3}$. At the experimental melting
temperature, the calculated density change remains essentially
unchanged, and the densities in both phases deviate only by $2$\%
with respect to the measurements.

\subsection{Liquid-solid interfaces}
\label{SubSec:MeltingCurve}

Liquid-solid interface (LSI) simulations are
performed for the purpose of determining the melting line by the
two-phase coexistence. The procedure follows the approaches proposed
in Refs.~\cite{Sun2004, Morris1994, Morris2002, Zykova2009, Zykova2010,
Kuhn2013, Benjamin2015, Rozas2016, Rozas2021} and is similar to the
protocol used in Ref.~\cite{Becker2020}. A simulation cell containing
around $N=43000$ atoms is set up with an initial crystalline
configuration with a shape corresponding to $28\times7\times7$ primitive cells on which
PBC are applied to the three directions of space. Starting at zero
pressure, this system is heated and equilibrated at constant pressure
to a temperature of $50$\,K below a guess of the melting temperature.
Half of the simulation cell in the $x$ direction is further heated
and maintained at a much higher temperature until a
complete melting is observed. The liquid part is then cooled down
and equilibrated at a temperature $50$\,K above the guess, thus
creating a solid-liquid coexistence containing two crystal-melt
interfaces due to the PBC. The simulation of the entire system is
pursued in the isobaric-isoenthalpic ensemble so that the temperature
of the LSI is an internal parameter free to evolve toward a steady
state corresponding to the thermodynamic melting temperature if
both phases survive. The simulation is continued for $1$~ns, and
the average melting temperature is determined on the last $100$~ps
when a steady position of the two interfaces is observed. If a
complete melting or solidification occurs, the procedure is started
over again with a refined guess of the melting temperature. This
procedure is repeated with subsequent higher pressures by first
shrinking the volume of the whole simulation cell from the coexistence
configuration at the preceding pressure and then increasing the
temperature at constant pressure to a new guess of the melting line.

\begin{figure}[t]
\centering
\includegraphics[scale=0.65]{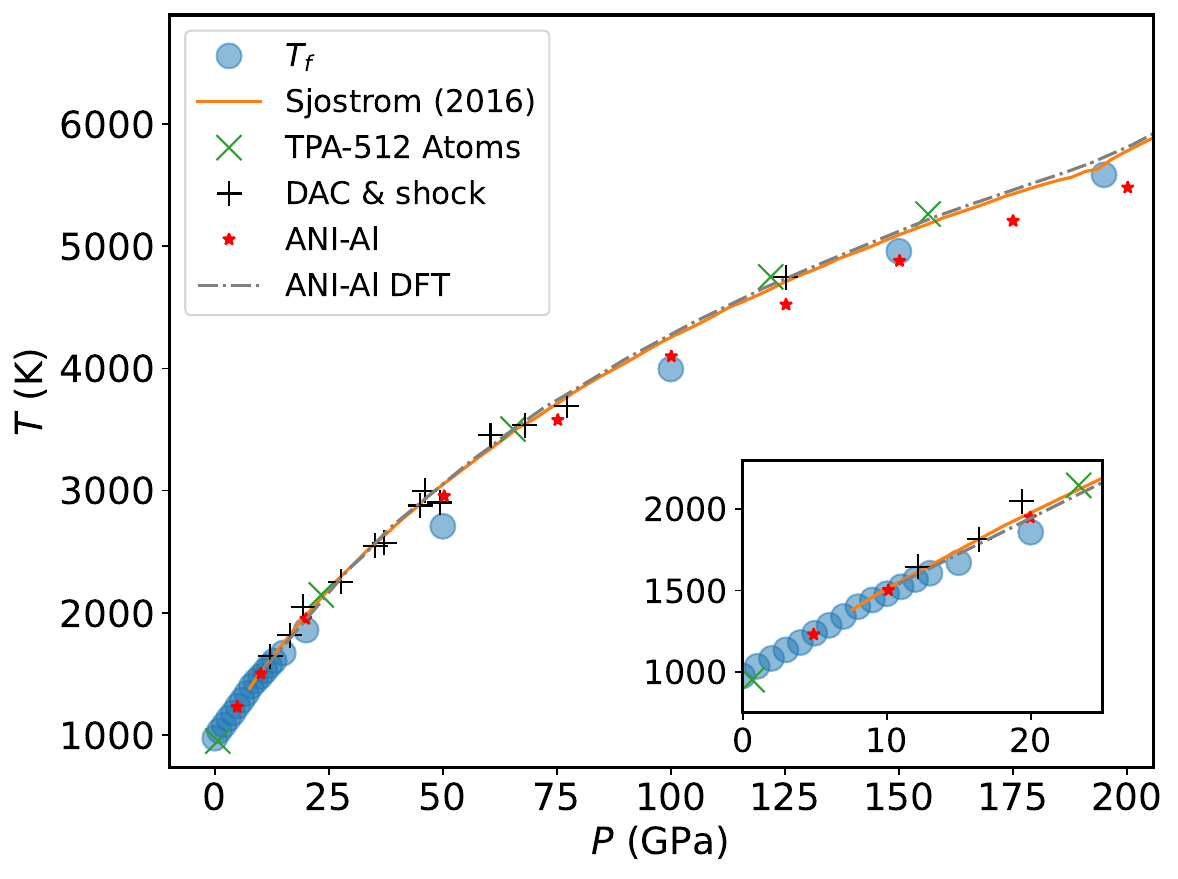}
\caption{\label{fig:M}
Melting curve for aluminium as a function of pressure. The blue circles obtained from our LSI simulations with the HDNN are compared to the experimental
data \cite{Boehler1997,Hanstrom2000,Shaner1984}, the AIMD two-phase
approach \cite{Bouchet2009}, and the ANI-Al ML potential \cite{Smith2021}. The Inset highlights the lower pressure range up to $25$\,GPa.}
\end{figure}

The melting curve of aluminium was measured \cite{Boehler1997,Hanstrom2000}
up to $80$~GPa using diamond anvil cells (DAC) and even higher at
$125$\,GPa by means of shock experiments \cite{Shaner1984}. In Fig.~\ref{fig:M}, 
the results obtained from the HDNN are compared to
these experimental data \cite{Boehler1997,Hanstrom2000,Shaner1984},
the \textit{ab initio} based equation of states (EOS) \cite{Sjostrom2016}
as well as the AIMD two-phase approach \cite{Bouchet2009} for which
the GGA for the XC functional and $512$ atoms were used. At ambient
pressure, the HDNN potential yields a value of $T_M=970$~K which
overestimates the experimental one of $933$\,K by 5\%. This is also
the case for the AIMD \cite{Bouchet2009} to a lesser extent,
recalling that the GGA was used and overestimates the atomic volume
\cite{Alfe2003,Bouchet2009}. With increasing pressure, the melting
curve from the HDNN potential slightly underestimates the experiments
as well as the EOS. By using two different sizes and shapes,
negligible influence on the determination of the melting curve was
found, confirming earlier results on pure Zr \cite{Becker2020}.
Noticeably, the reliability of the present potential on the melting
line up to $200$\,GPa is then assessed, even if high pressure thermodynamic states included in the training set are really scarce. { Interestingly, the HDNN curve is similar to the one obtained with the ANI-Al ML potential by Smith \textit{et al.} \cite{Smith2021}, and even gives slightly better results at high pressure.}

\subsection{Homogeneous Nucleation}
\label{SubSec:Nucleation}
\begin{figure}[t]
\centering
	
\includegraphics[scale=0.53]{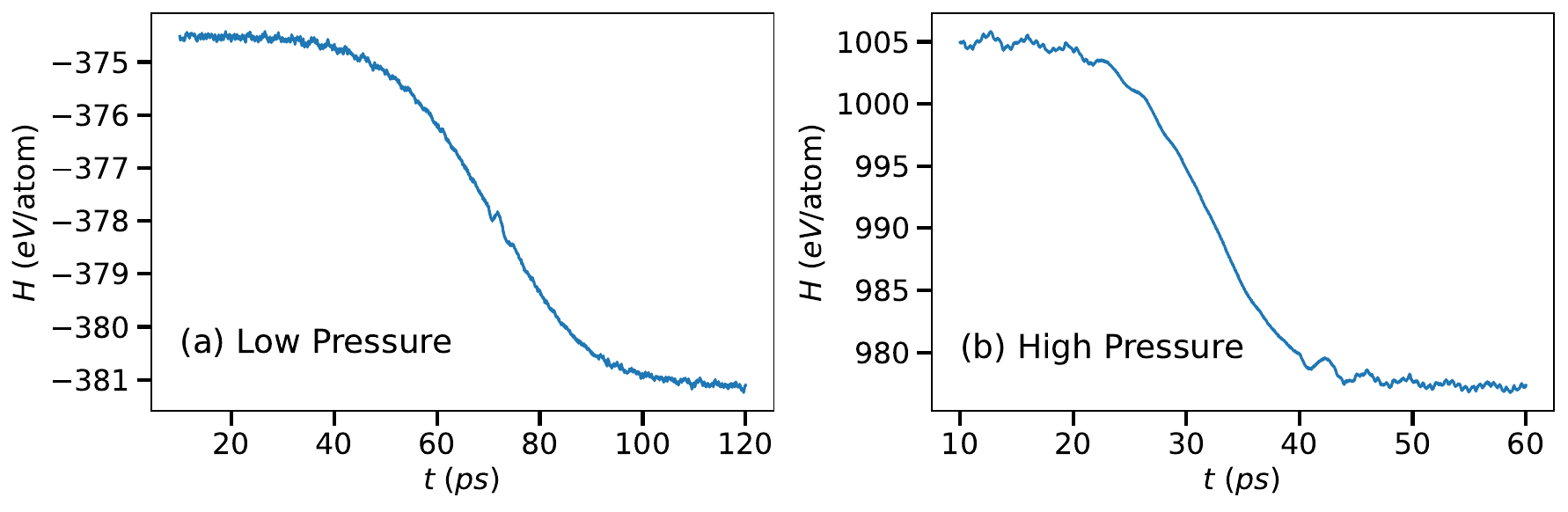}

\vcenteredhbox{\begin{minipage}{0.23\textwidth}
{\large \textbf{(c) $t = 30$~ps}}
\end{minipage}}	
\vcenteredhbox{\begin{minipage}{0.23\textwidth}
{\large \textbf{(d) $t = 40$~ps}}
\end{minipage}}
\vcenteredhbox{\begin{minipage}{0.23\textwidth}
{\large \textbf{(e) $t = 50$~ps}}
\end{minipage}}
\vcenteredhbox{\begin{minipage}{0.23\textwidth}
{\large \textbf{(f) $t = 60$~ps}}
\end{minipage}}

\includegraphics[scale=0.19]{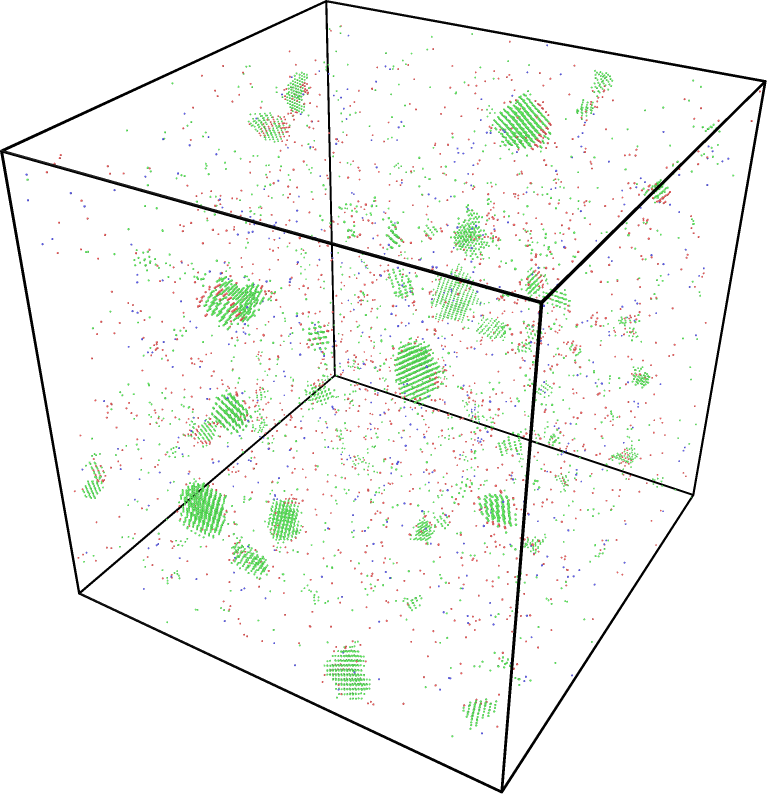}
\includegraphics[scale=0.19]{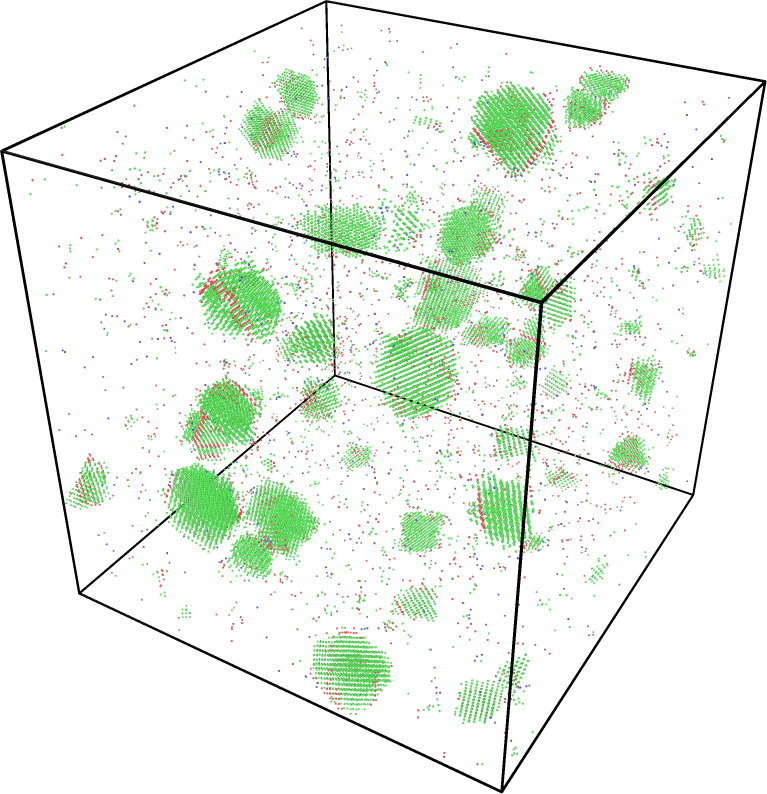}
\includegraphics[scale=0.19]{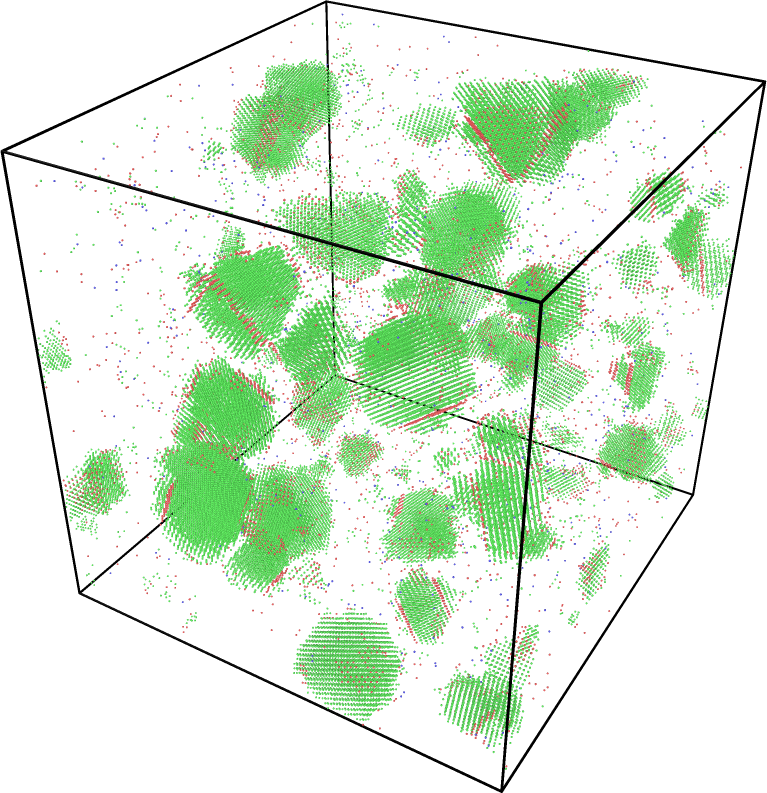}
\includegraphics[scale=0.19]{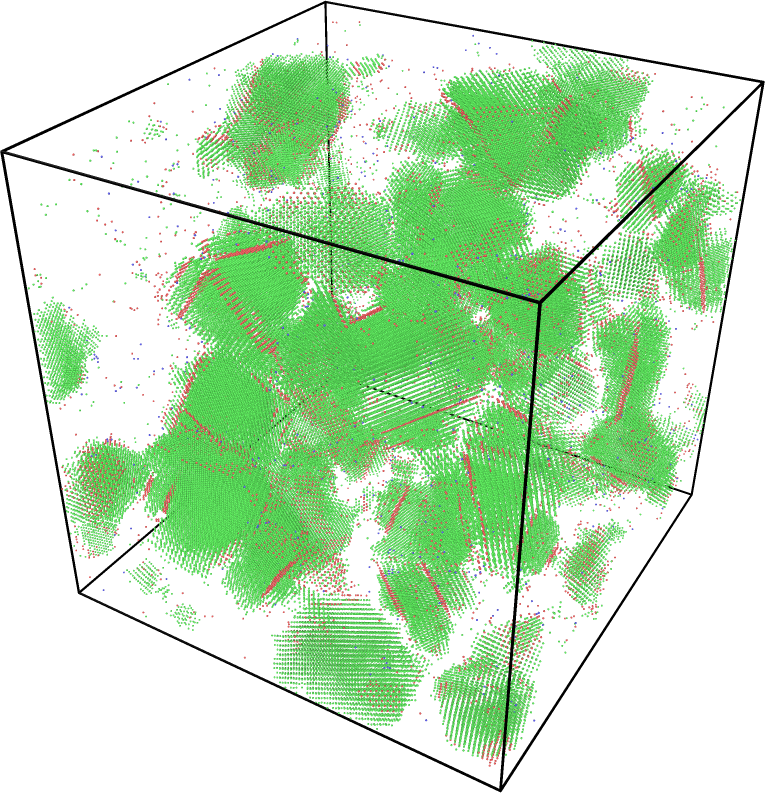}

\vcenteredhbox{\begin{minipage}{0.23\textwidth}
		{\large \textbf{(g) $t = 14$~ps}}
\end{minipage}}	
\vcenteredhbox{\begin{minipage}{0.23\textwidth}
		{\large \textbf{(h) $t = 20$~ps}}
\end{minipage}}
\vcenteredhbox{\begin{minipage}{0.23\textwidth}
		{\large \textbf{(i) $t = 26$~ps}}
\end{minipage}}
\vcenteredhbox{\begin{minipage}{0.23\textwidth}
		{\large \textbf{(j) $t = 32$~ps}}
\end{minipage}}

\includegraphics[scale=0.19]{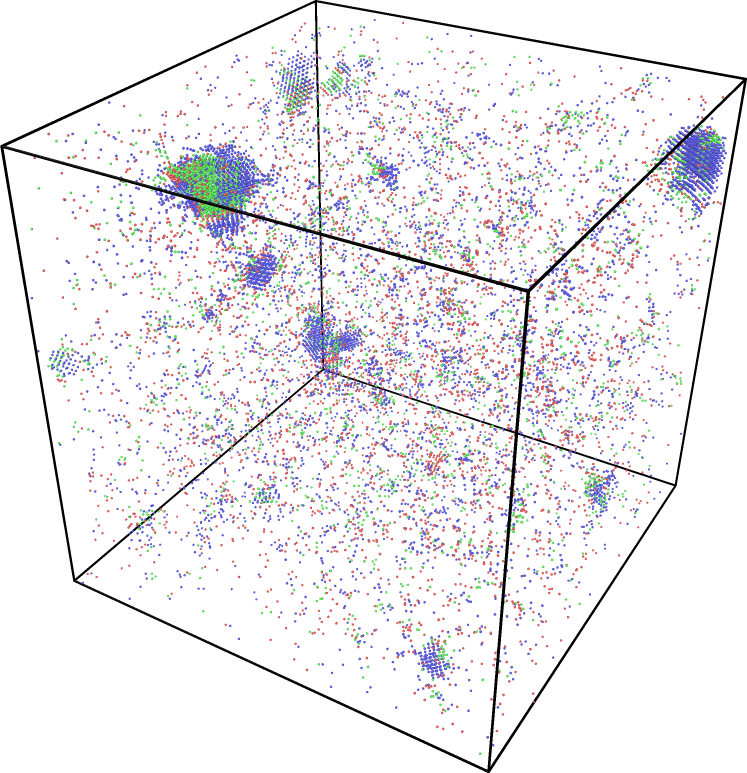} 
\includegraphics[scale=0.19]{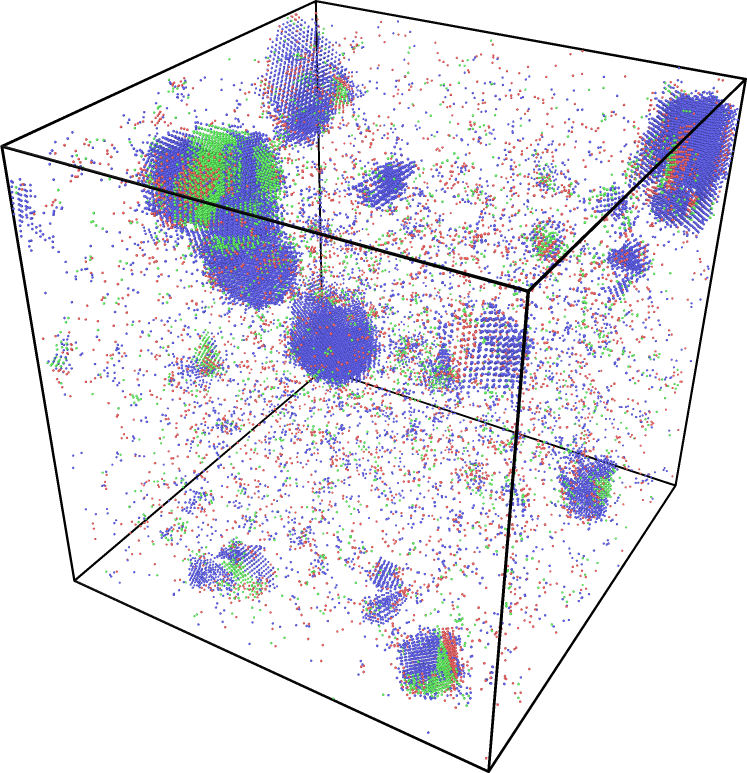} 
\includegraphics[scale=0.19]{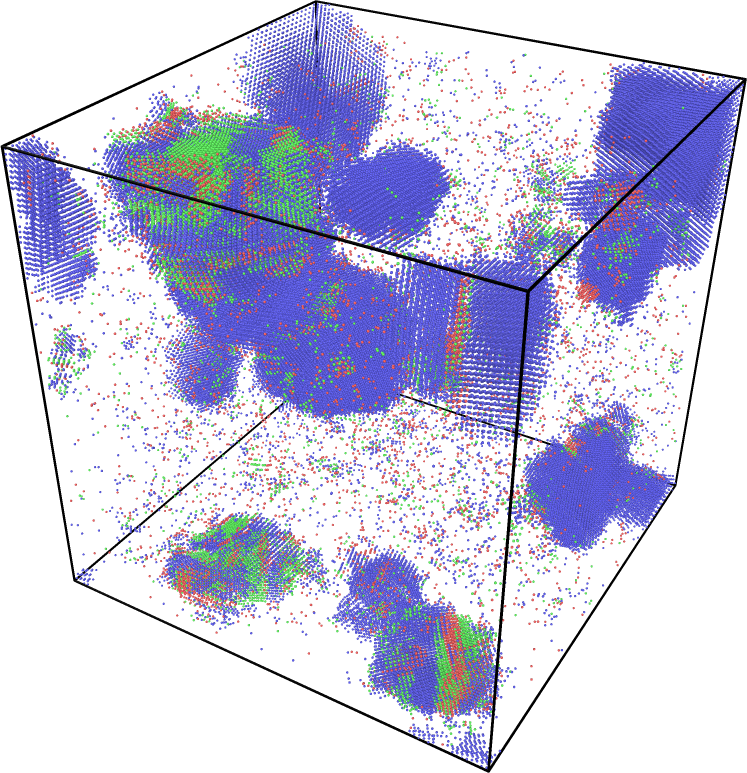} 
\includegraphics[scale=0.19]{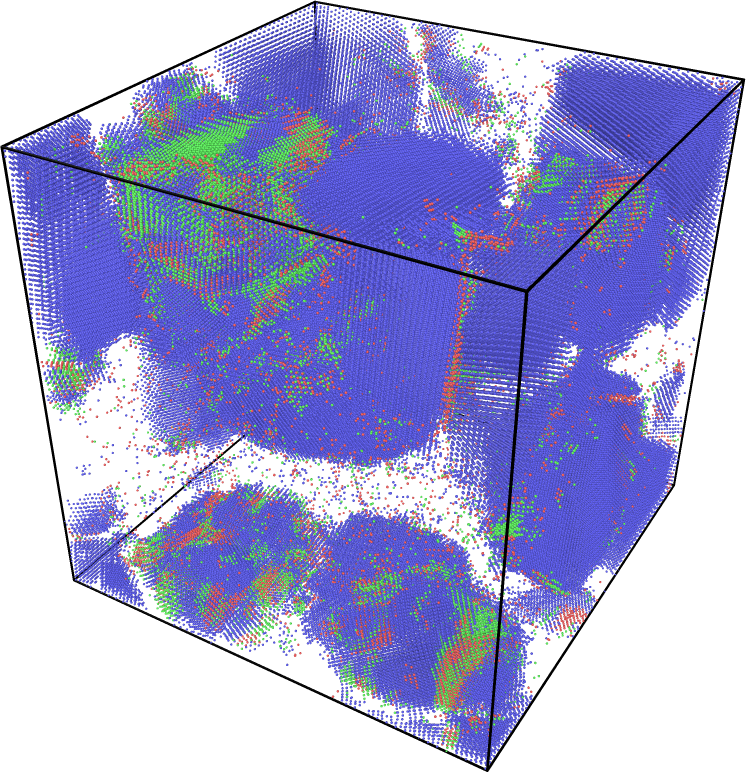} 
\caption{\label{fig:Homo}
Homogeneous nucleation of deeply undercooled aluminium along the
$T=600$\,K and $4000$\,K isotherms, respectively for ambient pressure and $200$\,GPa.
Time evolution of the enthalpy (a) at ambient pressure, and (b) at $200$\,GPa.
Snapshots of the simulation as various times during the nucleation (c) to (f) at ambient pressure, and (g) to (j) at $200$\,GPa.
Only the atoms with crystalline ordering in the sense of the
common-neighbor analysis are drawn: fcc (green); hcp (red); bcc
(blue).}
\end{figure}

Finally, the homogeneous nucleation is investigated and depicted
in Fig.~\ref{fig:Homo}. As pointed out in our preceding in our preceding work \cite{Becker2020}, a more accurate investigation require the use of large enough simulation boxes, with typically 1 million atoms or more. This allows the occurrence of multiple nuclei during the nucleation process. Therefore, the system of 1000188 atoms at ambient pressure is analyzed further along the $600$~K isotherm used for the determination of the TTT curve in ~\ref{fig:H}(b). At such a high degree of undercooling $\Delta T = (T_M-T)/T_M =0.38$,
an extremely fast nucleation process is observed \cite{Orava2014,Herlach2015}. Similarly, such a fast homogeneous nucleation is seen at the high pressure of $200$\,GPa along the  $4000$\,K with $\Delta T=0.24$. Inherent structure configurations shown in Figs. (c)-(j) were first analyzed using the common-neighbor analysis \cite{Faken1994} and only atoms with a crystalline environment (fcc, hcp, and bcc) are shown. As expected, nucleation occurs showing growing nuclei in the fcc ordering with hcp stacking faults at ambient pressure. At 200 GPa, nucleation starts with nuclei having a bcc order with sometimes some fcc ordering at their boundary that transform back to the bcc structure during the growth. These nucleation pathways are pretty much consistent with the ($P,T$) phase diagram \cite{Sjostrom2016} showing the reliability of the HDNN potential designed here.     

\begin{figure}[t]
  \centering
  \includegraphics[width=0.3\textwidth]{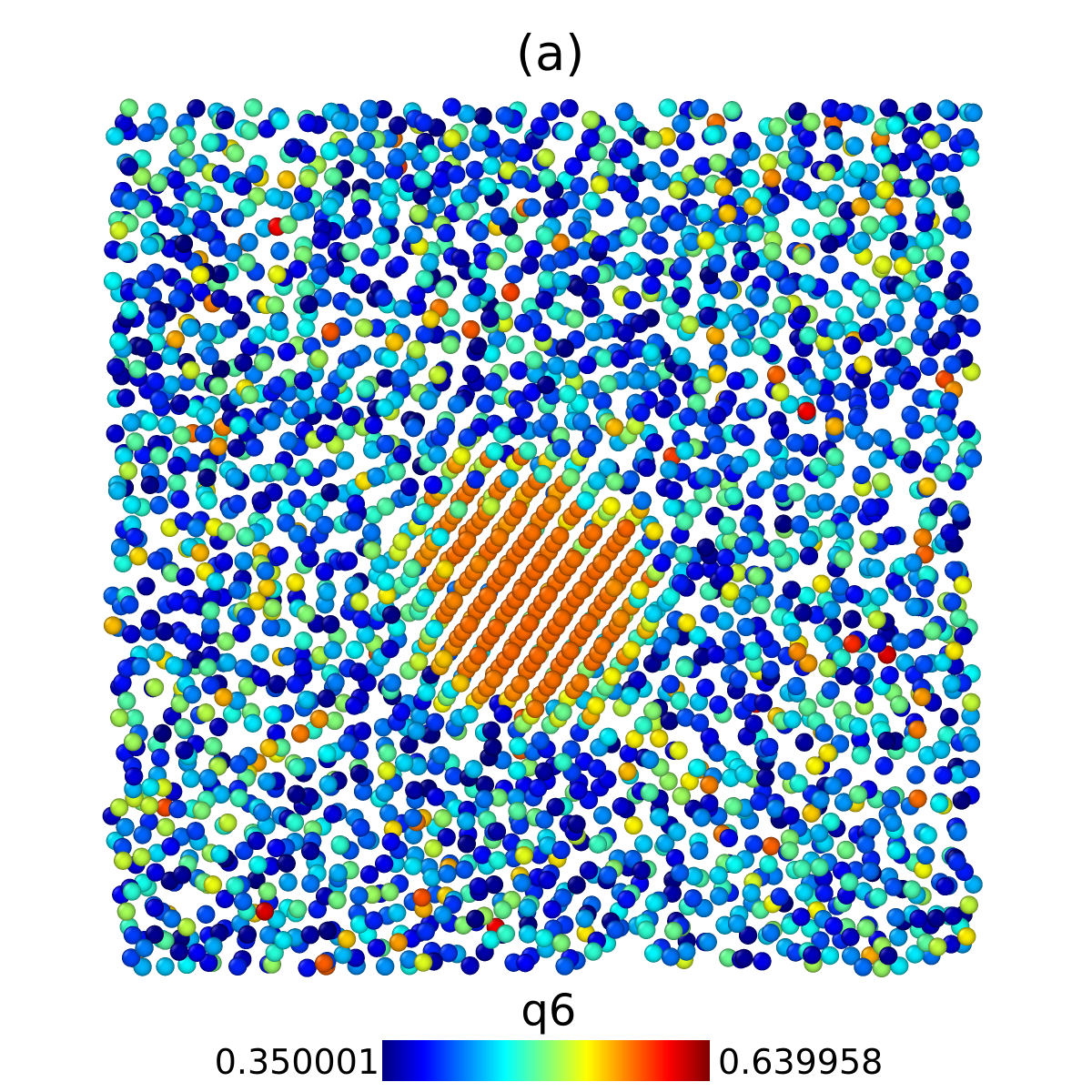}
  \includegraphics[width=0.33\textwidth]{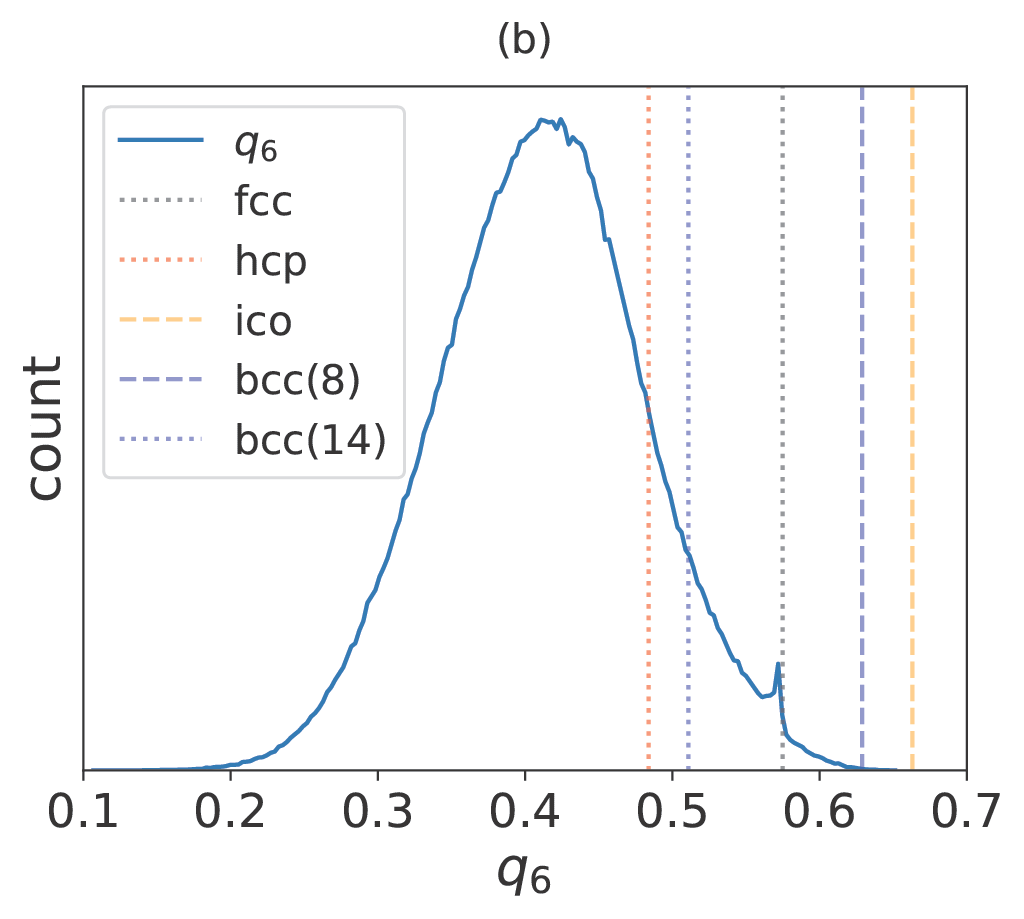}
  \includegraphics[width=0.33\textwidth]{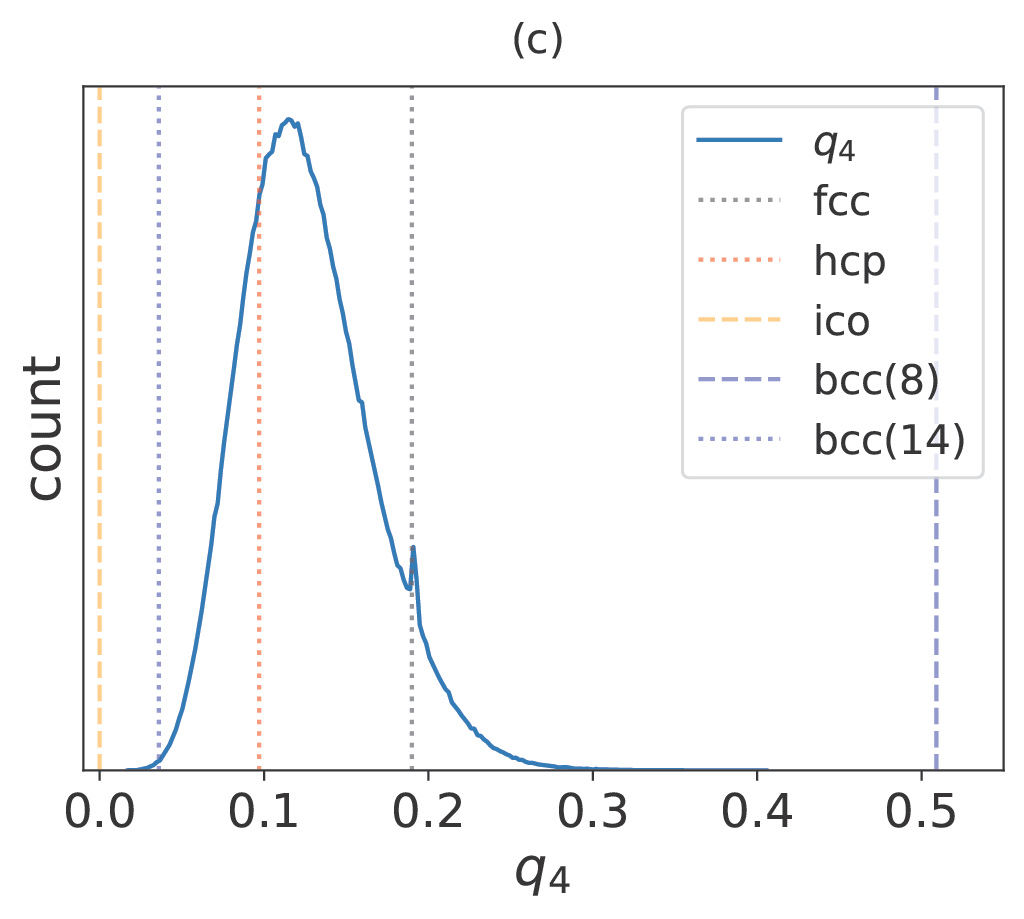}

  \caption{\label{fig:q6}
    {(a) First identified  nucleus at the onset of nucleation in a small box extracted from the snapshot at $30$\,ps (Fig. \ref{fig:Homo}) for ambient pressure. Atoms are colored according to he value of the $q_6$ averaged Steinhardt order parameter. (b-c) Corresponding histograms for all atoms in the simulation box of the $q_6$ and $q_4$ parameters showing clearly that nucleation starts with only a fcc ordering.}}
\end{figure}

{ The onset of nucleation occurs at about $30$ ps at ambient pressure and $4$\,ps at $200$\,GPa, earlier that the nucleation time defined for the construction of the TTT curve above.  Using the averaged Steinhardt Order Parameters $q_4$ and $q_6$ \cite{Lechner2008}, embryos with atoms having a crystalline ordering showed that they dissolve back to the liquid with a size less than $90$ atoms in both cases. The latter value
does not represent \textit{per se} the size critical nucleus but rather a lower limit. As expected at ambient pressure, the main crystalline phase during the growth was identified as fcc, as can be seen in Fig.~\ref{fig:q6}(a), but a significant hcp ordering also appears during the nucleation
and remains as stacking fault after complete solidification of the simulation box as shown in Fig. \ref{fig:Homo}. At high pressure, the onset of nucleation occurs in the bcc ordering, with fcc and hcp ordering at their surface at later stages.}

It is an open question in general, whether the homogeneous nucleation
process follows the Landau Theory in which the bcc precursor is
favored in the early stages of crystal nucleation \cite{Alexander1974}
or the Ostwald step rule \cite{Ostwald1897} according to which a
primary crystal phase could be different from the the fcc one. { From the distributions of the $q_6$ and $q_4$ shown in Figs. \ref{fig:q6}(b) and \ref{fig:q6}(c) only fcc ordering emerges at the onset of nucleation.} Our
findings show that aluminium follows a single step process with an
onset of homogeneous nucleation showing emerging embryos with a fcc
ordering. The resulting nuclei grow in a rather patchy shape with
a small amount of hcp stacking fault defects. This nucleation the
scenario is different from the Lennard-Jones case
\cite{ten1995,tenWolde1996} which follows the Landau theory and the
Ostwald step rule. The present {large-scale molecular dynamics results with close ab initio accuracy allows us to assess very recent molecular
dynamics simulations \cite{Mahata2018,Becker2021} with EAM empirical potentials}. Our findings further show that such a single step nucleation pathway also occurs at high pressure with bcc ordering in the emerging nuclei.  

\section{Conclusion}
\label{Sec:Conclusion}
In the present work, a machine learning potential for
pure aluminium by means of a high dimensional neural network on the
basis of the well-known Behler-Parrinello approach \cite{Behler2007,Singraber2019b} was developed. This ML potential is devoted to the description of condensed phases,
namely liquid and solid states at ambient pressure as well as those
at pressures up to $300$\,GPa with resulting temperatures as high
as $8000$\,K. A crucial point was the training of the potential with
a data set generated by DFT-based simulations not only to cover the
targeted domain of thermodynamic states for a question of transferability
but also to consider for each of them in a physical meaningful
manner the time fluctuations by an appropriate sampling of phase
space trajectories obtained by \textit{ab initio} molecular dynamics.
This allows to include in the training the relevant accessible
microstates of the considered thermodynamic states. Another approach
based on metadynamics was shown to be efficient in selecting the
relevant configurations to train the neural network \cite{Bonati2018}.

The HDNN potential thus obtained was shown to be efficient in
reproducing the structural, dynamics as well as thermodynamic
quantities in the liquid, undercooled and crystalline states at
ambient pressures as well as in the liquid state at high pressure
up to $300$\,GPa, including the melting line. { One important outcome
is that a reliable ML potential could be obtained without including
explicitly the forces in the training  by using an appropriate sampling of AIMD trajectories. The procedure was shown} for Al and HDNN to perform well, giving a RMSE on forces similar to what is current obtained. The early stages of the homogeneous crystal nucleation was further investigated
on a scale much larger than what is possible from the \textit{ab
initio} molecular dynamics but with a similar accuracy. Results
show that aluminium follows a single step nucleation process with
an emerging fcc ordering and hcp stacking fault defects, confirming
recent works using large scale molecular dynamics
\cite{Mahata2018,Becker2021}, and also consistent with very recent simulations on nucleation during cooling \cite{Zhou2021}. A single step nucleation pathway with bcc nuclei is also observed at high pressure. 

Finally, the fact that the HDNN potential keeps a good accuracy
even in domains where the thermodynamic states in the training set
are scarce opens up a research line based upon active learning for
regression approaches to reduce efficiently the training set.
Dynamical properties such as the diffusion coefficients considered
here are sensitive to the details of the potential and should be
introduced in the training procedure in a more direct way than
through the choice of the XC functional in the DFT calculations.
This would represent a real step forward in designing ML potentials.
\section*{Acknowledgments}
We acknowledge the CINES and IDRIS under Project No. INP2227/72914,
as well as CIMENT/GRICAD for computational resources. This work was
performed within the framework of the Centre of Excellence of
Multifunctional Architectured Materials “CEMAM” ANR-10-LABX-44-01
funded by the “Investments for the Future” Program. This work has
been partially supported by MIAI@Grenoble Alpes (ANR-19-P3IA-0003).
Fruitful discussions within the French collaborative networks in high-temperature thermodynamics GDR CNRS 3584
(TherMatHT) and in artificial intelligence in materials science GDR CNRS 2123 (IAMAT) are also acknowledged. We thank J. Smith and K. Barros for their kind help in setting up the simulations with their ANI-Al potential from Ref. \cite{Smith2021}.
J.~S.\ acknowledges funding from the German Academic Exchange Service 
(DAAD) through the DLR-DAAD programme, grant No.~509.

\renewcommand{\thefigure}{S\arabic{figure}}
\renewcommand{\thetable}{S\Roman{table}}

\begin{center}
	{\Large Supplementary Information File}
\end{center}
\maketitle

\section{Dataset for the training of the HDNN potential}
Tables \ref{tab:training} and \ref{tab:training2} gather all thermodynamics states that have been simulated by AIMD to generates configuration for the training procedure of the HDNN potential. For each state, after an equilibration at a target temperature a phase space trajectory was produced from which a sample of configurations was randomly extracted to include in the data set. Additional AIMD at $T=950$ K and $P=0$ GPa, $1500$ K and $2.5$ GPa,  $2000$ K and $28$ GPa, and $8000$ K and $227$ GPa were performed but were not included in the data set for the sake of testing the predictive ability of the HDNN potential. The database contains in total $24300$, which amounts to $6$ $220$ $800$ atoms.  
\begin{table}[t]
	\begin{center}
		\begin{tabular}{cccccc}
			\hline\hline
			\makebox[2.6cm][c]{Structure} & \makebox[2.6cm][c]{$T$ (K)} & \makebox[2.6cm][c]{$P$ (GPa)} & \makebox[2.6cm][c]{Trajectory (ps)} & \makebox[2.6cm][c]{Sample size} & \makebox[2.6cm][c]{ } \\
			\hline
			fcc & $10$ & $0$ & $40$ & $1000$&\\
			fcc & $300$ & $0$ & $40$ & $1000$&\\
			fcc & $400$ & $0$ & $40$ & $1000$&\\
			fcc & $500$ & $0$ & $40$ & $1000$&\\
			fcc & $600$ & $0$ & $40$ & $1000$&\\
			fcc & $700$ & $0$ & $40$ & $1000$&\\
			fcc & $800$ & $0$ & $40$ & $1000$&\\
			fcc & $10$ & $1$ & $10$ & $100$&\\
			fcc & $10$ & $10$ & $10$ & $100$&\\
			fcc & $10$ & $100$ & $10$ & $100$&\\
			hcp & $10$ & $0$ & $10$ & $100$&\\
			hcp & $10$ & $10$ & $10$ & $100$&\\
			hcp & $10$ & $100$ & $10$ & $100$&\\
			hcp & $10$ & $200$ & $10$ & $100$&\\
			hcp & $10$ & $300$ & $10$ & $100$&\\
			bcc & $10$ & $0$ & $10$ & $100$&\\
			bcc & $10$ & $10$ & $10$ & $100$&\\
			bcc & $10$ & $100$ & $10$ & $100$&\\
			bcc & $10$ & $200$ & $10$ & $100$&\\
			bcc & $10$ & $300$ & $10$ & $100$&\\
			\hline\hline
		\end{tabular}
	\end{center}
	\caption{\label{tab:training} Characteristics of the data set built from AIMD simulations. Are given the structure of the simulation (fcc, hcp, bcc, and liquid), the temperatiure $T$, the pressure $P$, the time span of the AIMD trajectory from which the configuration are sampled, the sample size, namely the number of configurations randomly extracted from the trajectory. The pressures less that $1$ GPa were indicated as $0$.}
	
\end{table}
\begin{table}[t]
	\begin{center}
		\begin{tabular}{cccccc}
			\hline\hline
			\makebox[2.6cm][c]{Structure} & \makebox[2.6cm][c]{$T$ (K)} & \makebox[2.6cm][c]{$P$ (GPa)} & \makebox[2.6cm][c]{Trajectory (ps)} & \makebox[2.6cm][c]{Sample size} & \makebox[2.6cm][c]{ } \\
			\hline
			
			Liquid & $500$ & $0$ & $40$ & $1000$&\\
			Liquid & $600$ & $0$ & $40$ & $1000$&\\
			Liquid & $650$ & $0$ & $40$ & $1000$&\\
			Liquid & $700$ & $0$ & $40$ & $1000$&\\
			Liquid & $750$ & $0$ & $40$ & $1000$&\\
			Liquid & $800$ & $0$ & $40$ & $1000$&\\
			Liquid & $1000$ & $0$ & $40$ & $1000$&\\
			Liquid & $1100$ & $0$ & $40$ & $1000$&\\
			liquid & $1250$ & $0$ & $40$ & $1000$&\\
			liquid & $1350$ & $0$ & $40$ & $1000$&\\
			liquid & $1500$ & $0$ & $40$ & $1000$&\\
			liquid & $1600$ & $0$ & $40$ & $1000$&\\
			liquid & $1700$ & $0$ & $40$ & $1000$&\\
			liquid & $3100$ & $56$ & $40$ & $1000$&\\
			liquid & $4500$ & $107$ & $40$ & $1000$&\\
			liquid & $8000$ & $320$ & $40$ & $1000$&\\
			
			\hline\hline
		\end{tabular}
	\end{center}
	\caption{\label{tab:training2} Characteristics of the data set built from AIMD simulations (ontinued). Same caption as Table \ref{tab:training}.}
	
\end{table}

\section{EAM and MEAM pair-correlation functions}

\begin{figure}
	\centering
	\includegraphics[width=0.49\textwidth]{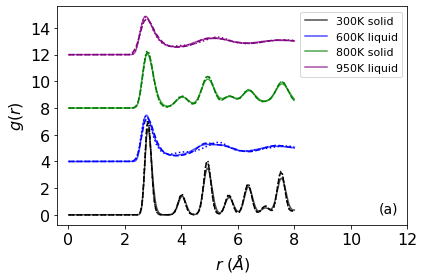}
	\includegraphics[width=0.49\textwidth]{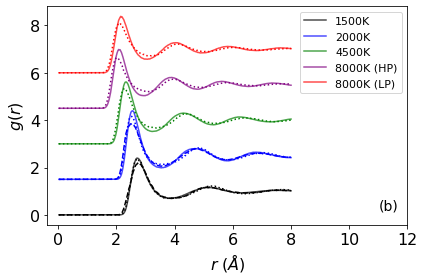}
	\caption{\label{fig:eam-meam}  Pair-correlation functions obtained using the Embedded Atom Model (dashed lines) and Modified Embedded Atom Model (dotted lines) potentials, with solid lines showing the corresponding g(r) obtained from AIMD simulations. Lines of same color correspond to the same temperature and system volume. The plots for $T=600$, $800$, and $950$ K are shifted upwards by $4$, $8$, and $12$. Likewise the plots for $T=2000$, $4500$, $8000$ K (HP), and $8000$ K (LP) are shifted by $1.5$, $3$, $4.5$, and $6$. The two $8000$ K lines correspond to high ($360$ GPa) and low ($250$ GPa) pressure.}
\end{figure}

\begin{table}[t]
	\begin{center}
		\begin{tabular}{cccccc}
			\hline\hline
			\makebox[2.6cm][c]{$T$ (K)} & \makebox[2.6cm][c]{$300$} & \makebox[2.6cm][c]{$600$} & \makebox[2.6cm][c]{$800$} & \makebox[2.6cm][c]{$950$} & \makebox[2.6cm][c]{ }\\
			\hline
			MSE (EAM) & $0.0297$ & $0.0041$ & $0.0048$ & $0.0032$&\\
			MSE (MEAM) & $0.036$ & $0.0065$ & $0.0064$ & $0.0018$&\\
			$p$ (EAM) & $5.49\times10^{-6}$ & $3.59\times10^{-4}$ & $4.48\times10^{-6}$ & $3.71\times10^{-6}$&\\
			$p$ (MEAM) & $5.48\times10^{-7}$ & $2.20\times10^{-13}$ & $3.98\times10^{-9}$ & $3.78\times10^{-10}$&\\
			\hline\hline
			\\
			\hline\hline		
			$T (K)$ &$1500$& $2000$& $4500$& $8000$ (H) & $8000$ (L) \\
			\hline
			MSE (EAM)  & $0.0028$ & $0.012$ & N/A & N/A & N/A \\
			MSE (MEAM) & $0.00086$ & $0.066$ & $0.0091$ & $0.015$ & $0.0095$ \\
			$p$ (EAM)  & $4.24\times10^{-5}$ & $1.51\times10^{-6}$ & N/A & N/A & N/A \\
			$p$ (MEAM)  & $5.96\times10^{-7}$ & $9.63\times10^{-16}$ & $1.04\times10^{-11}$ & $7.44\times10^{-11}$ & $5.92\times10^{-11}$\\
			\hline\hline
		\end{tabular}
	\end{center}
	\caption{\label{tab:eam-meam}	Mean-square error of the pair-correlation function $g(r)$ obtained \textit{via} the EAM and MEAM potentials, and evaluated against the one obtained from AIMD, with corresponding $p$-values  (noted here as $p$) as explained in the text.}
	
\end{table}

To get a sense of how our HDNNP compares to other widely used potentials, we have run simulations using the Embedded Atom Model (EAM) \cite{Medelev2008} and Modified Embedded Atom Model (MEAM) \cite{Lee2003} potentials both known as performing well in the liquid and solid states (see Ref. \cite{Mahata2018} and references therein).
These simulations were performed identically to the ones used to obtain the pair-correlation functions for our HDNNP, and for the same set of temperatures and densities (see the main text). Figure \ref{fig:eam-meam} shows g(r) obtained from these simulations, along with the corresponding AIMD ones. For the sake of clarity, the results of the HDNN potential are not shown since they match very closely the AIMD curves. It is worth mentioning that above $20$ GPa we were not able to achieve MD simulations with the EAM potential. At ambient pressure, both EAM and MEAM reproduce well the AIMD simulations even if noticeable can bee seen. At larger pressures and temperature the agreement worsen showing that the EAM is not transferable while the MEAM still gives reasonable results. 

he mean square error between the two were calculated, as shown in table \ref{tab:eam-meam}. In this table, are given the  $p$-values obtained by performing a $t$-test statistics between the square errors $(g_\mathrm{AIMD}(r)-g_\mathrm{NNP}(r))^2$ and $(g_\mathrm{AIMD}(r)-g_\mathrm{emp.}(r))^2$, treating the error at different radii as independent.

\section{Steinhardt parameter analysis}
{
	To study the local ordering before and during nucleation we have used the Steinhardt bond-ordering parameters\cite{Steinhardt1983}, more specifically the averaged form\cite{Dellago2008}, as implemented in the pyscal code\cite{pyscal2019}.
	First define for each atom $i$ the vector
	\begin{equation}
	q_{lm}(i) = \frac{1}{N(i)} \sum_{j=1}^{N(i)} Y_{lm}(\mathbf{r}_{ij})
	\end{equation}
	where $N(i)$ is the number of nearest neighbors of atom $i$, $\mathbf{r}_{ij}$ is the displacement of nearest-neighbor atom $j$ from $i$, and $Y_{lm}$ is the spherical harmonics.
	From these the averaged bond-order parameters can be defined as
	\begin{equation}
	\bar{q}_{l}(i) = \sqrt{\frac{4\pi}{2l + 1}\sum_{m=-l}^{l}
		\left|\frac{1}{N(i)+1}\sum_{k=0}^{N(i)}q_{lm}(k)\right|^2}
	\end{equation}
	where the sum from $k=0$ to $N(i)$ includes both the atom $i$ and its nearest neighbors.
	Due to being averaged over nearest neighbors, these paramers take into account not just the first coordination shell, but also the second.
	To perform structural analysis using these parameters one typically selects specific values for $l$, with $l=4$ and $l=6$ being a common choice.
	It is then possible to compare the resulting values of $\bar{q}_l$ to those of ideal crystals in order to identify crystal structure.

}

\end{document}